\documentclass[10pt]{article}
\usepackage{amssymb,bbm,amsmath,stmaryrd,subfig,color,mathrsfs,graphicx,pxfonts}
\usepackage[paperwidth=16cm,paperheight=24cm,textwidth=10cm,textheight=10cm,top=2cm,left=2cm,right=2cm,bottom=2cm]{geometry}
\usepackage{hyperref}
\definecolor{refcolor}{rgb}{0.3,0.3,0.7}
\hypersetup{
    pdftitle={From infinitesimal to full contact between rough surfaces: evolution of the contact area},
    pdfauthor={Vladislav A. Yastrebov},
    pdfsubject={Computational contact mechanics},
    pdfcreator={LaTeX, Kile},
    pdfkeywords={contact mechanics, rough contact, surface roughness, elastic contact, contact area, true area, error estimation, boundary element method},
    pdfnewwindow=true,
    colorlinks=true,
    linkcolor=refcolor,          
    citecolor=refcolor,          
    filecolor=refcolor,      
    urlcolor=refcolor           
}

\newcommand{\be}{\begin{equation}} \newcommand{\ee}{\end{equation}}
\renewcommand{\ee}{\end{equation}}
\newcommand{\bes}{\begin{equation*}}
\newcommand{\ees}{\end{equation*}}

\newcommand{\ddp}[2]{\frac{\partial{#1}}{\partial{#2}}}

 \usepackage{xspace}

 \renewcommand{\be}{\begin{equation}}

\begin{document}

 \title{From infinitesimal to full contact between rough surfaces: evolution of the contact area}
 
 \author{Vladislav A. Yastrebov$^a$ \qquad Guillaume Anciaux$^b$\\Jean-Fran\c cois Molinari$^b$}

 \date{\footnotesize$^a${\it Centre des Mat\'eriaux, MINES ParisTech, CNRS UMR 7633, BP 87, F 91003 Evry, France}\\
                    $^b${\it Computational Solid Mechanics Laboratory (LSMS, IIC-ENAC, IMX-STI), Ecole Polytechnique
F\'ed\'erale de Lausanne (EPFL), B\^at. GC, Station 18, CH 1015 Lausanne, Switzerland}
}
  
%
%
%

\maketitle

\begin{abstract}

We carry out a statistically meaningful study on self-affine rough surfaces in elastic frictionless non-adhesive contact.  
We study the evolution of the true contact area under increasing squeezing pressure.
Rough surfaces are squeezed from zero up to full contact, which enables us to compare the numerical results
both with asperity based models at light pressures and with Persson's contact model for the entire range of pressures.
Through the contact perimeter we estimate the error bounds inherent to contact area calculation in discrete problems.

A large number of roughness realizations enables us to compute reliably the
derivative of the contact area with respect to the pressure.
In contrast to Persson's model and in agreement with asperity based models, we demonstrate that at light pressures it is a decreasing convex function.
The nonlinearity of the contact area evolution, preserved for the entire range of pressures, is especially strong close to infinitesimal contact.
This fact makes difficult an accurate investigation of contact at light pressure and 
prevents the evaluation of the proportionality coefficient, which is predicted by analytical models.
A good agreement of numerical results with Persson's model is obtained for the shape of the area-pressure curve especially near full contact.
We investigate the effects of the lower and upper cutoff wavenumbers (longest and shortest wavelengths in surface spectrum, respectively), 
which control the Gaussianity of surface and spectrum breadth (Nayak's parameter), onto the contact area evolution.
While being one of the central characteristics of rough surfaces, the Nayak's parameter plays also quite an important role in rough contact mechanics,
but its effect is significantly weaker than predicted by asperity based models.
We give a detailed derivation of a new phenomenological contact evolution law; 
also we derive formulae that link Nayak's parameter and density of asperities with Hurst exponent and cutoffs wavenumbers.

\end{abstract}


\tableofcontents

\section{Introduction} 

Contact, adhesion and friction play an important role in many natural
(e.g., earthquakes) and engineering systems, for example, assembled parts in
engines, railroad contacts, bearings and gears, breaking systems,
tire-road contacts, metal forming, vehicle crash, bio mechanics,
granular materials, electric contacts, liquid sealing, etc. 
In all these examples, the contacting surfaces are rough.  
Being in dry contact (in absence of lubrication) means that the contacting solids touch each other at
many separate spots, whose area may be drastically different from the prediction of classical Hertz's contact theory. 
This roughness and complexity of the contact interface may be a major factor in analysis
of such systems for strength, critical stresses, fatigue and damage,
fracture initiation, friction, adhesion, wear, heat and electric charge transfer, and percolation. 
Real or true contact area is one of the central characteristics of the contact between rough surfaces.  
In this paper we analyze by means of numerical analysis how the real
contact area changes with applied pressure and what are the relevant
properties of the surface roughness that influence this evolution.
The numerical results are compared with existing analytical models and numerical results of
other authors.  We consider the problem of rough contact in its
simplest formulation: frictionless and non-adhesive contact between
linearly elastic half-spaces.  Regardless of the apparent simplicity of the
problem and multiple analytic/experimental/numerical studies, many questions remain open.

\subsection{Roughness}

All surfaces in nature and industry are rough under certain magnification.
This roughness possesses specific characteristics.
Most of rough surfaces are self-similar or self-affine, i.e. the roughness scales under magnification with a given
scaling coefficient all along the magnification range from macroscopic down to nanometric scales. 
Typical examples of this scaling are found in Earth landscapes, coast line, tectonic faults, ocean's surface and
engineering surfaces~\cite{thomas1999b,meakin1998b}. 
Among a wide variety of rough surfaces, the class of isotropic Gaussian surface
deserves a particular attention from the scientific community due to
its relative simplicity and generality~\cite{longuethiggins1957rsla,nayak1971tasme,greenwood1966prcl,bush1975w}.
By isotropy one implies that statistical properties of any two
profiles measured along different directions are identical.
By normality or Gaussianity of a surface one implies that surface heights are normally distributed.

The self-affinity of rough surfaces may be decoded by analysis of its
autocorrelation function $R(x,y)$ or the Fourier transform of $R$
which is called the power spectral density (PSD) $\Phi(k_x,k_y)$,
where $k_x,k_y$ are the wavenumbers\footnote{Hereinafter by a wavenumber we imply a spectroscopic wavenumber normalized by the sample length $L$ to render them dimensionless.} in orthogonal directions $x,y$.
For many natural and engineering surfaces, the PSD decays as a power-law of the wavenumber~\cite{majumdar1990w, dodds1973jsv, vallet2009ti}:
$$ \Phi(k_x,k_y) \sim \left[\sqrt{k_x^2+k_y^2}\right]^{-2(1+H)},
$$ where $H$ is the Hurst roughness exponent which is related to the
fractal dimension $D$ as $D=3-H$. 
The PSD is bounded at the upper scale by the longest wavelength $\lambda_l$ (or the smallest wavenumber $k_l=L/\lambda_l$).
To handle a continuum model of a rough surface, the PSD may be bounded at the lower scale by the shortest wavelength $\lambda_s$ (or the highest wavenumber $k_s=L/\lambda_s$) (for detailed discussion see Section~\ref{sec:roughness}). 

\subsection{\label{ssec:mechanics}Mechanics of rough contact}

The surface roughness has important consequences on the mechanics and physics of contact.
For instance, the widely used Hertz theory of contact~\cite{hertz82,johnson1987b}, in which the contacting surfaces are
assumed to be smooth, is not valid for rough surfaces,
as the roughness induces high fluctuations of local deformations close to the contact surface, that go easily beyond the elastic limits and/or
fracture strength of materials.
This fact follows directly from the observation, that for most materials and loads the real contact area $A$ between contacting solids is only a small
fraction of the nominal (apparent) contact area $A_0$ predicted by Hertz theory.
The real contact area characterizes the transfer of heat and electricity through the contact interface,
frictional properties of the contact as well as the strength of adhesion and amount of wear.

The stochastic nature of rough contact makes it difficult to estimate
material rupture or stick-slip transition within a deterministic
approach and requires a probabilistic description and a statistical
analysis.  Many factors affect the mechanics of rough contact.  For
example, the real contact area depends on mechanics (contact pressure, friction, adhesion, wear), 
on multi-physics effects (Joule heating in electric contact, chemical reactions, frictional heating), on
time (viscosity and aging of materials) and environment (oxidation of
surfaces, temperature, humidity). While in experiments it is hard to study
all these aspects separately and deduce the more relevant ones,
in numerical simulations it is difficult to include many mechanisms to study their combined effect, as 
the models become excessively complex and hardly verifiable. In experiments, the contacting surfaces are also
hard to observe \textit{in situ} to characterize directly the contact zones. Thus indirect observation methods were 
adapted (measurements of the heat and electric transfer through the contact interface~\cite{bowden2001b}), which may bias the measurements due to 
the strong coupling between involved phenomena. 

Another challenge in rough contact arises from the breakdown of
continuum contact mechanics at nano-scale~\cite{robbins2005n}.
This issue is relevant if the roughness is present at atomic scale~\cite{krim1995ijmpb}, which is often the
case~\cite{Misbah2010rmp,Einax2013rmp} particularly for crystalline
materials for which dislocations reaching free surfaces leave atomic ``steps'' on them.
This atomic roughness can be taken into account by means of atomic modeling~\cite{Sinnott2008b,Spijker2013ti}.
But it is particularly hard to link the atomistic simulations of rough surfaces with
macroscopic results as there is a lack of representativity in analyzed
samples.  As there is no scale separation in surface roughness,
classical hierarchical homogenization models cannot be directly applied to
the analysis of rough contact. However, coupling between atomistic simulations with
finite element models~\cite{ramisetti2013cmame,ramisetti2013ijnme} (eventually accompanied with discrete dislocation dynamics coupling) 
is a promising technique to perform large simulations of contact between rough surfaces at atomic scale~\cite{anciaux2009ijnme,anciaux2010ijnme}.

To remain in the framework of continuum mechanics, one needs to
abandon the atomistic scale and introduce an artificial short wavelength cutoff
$\lambda_s$ in the surface to obtain a roughness which is smooth under a certain magnification. 
Consideration of such surfaces with truncated fractality lies in the foundation of classical
analytical models of rough contact; moreover, valid numerical
studies are only possible on surfaces which are smooth enough.
Normally, at the longest wavelengths, real surfaces do not demonstrate
self-affinity and the PSD has a plateau for a certain range of
wavelengths~\cite{persson2005jpcm}. 
This plateau includes wavelengths from $\lambda_l$ to $\lambda_r$, where $\lambda_l$ is the longest wavelength and $\lambda_r$ is a so-called rolloff wavelength.
So the fractality of rough surfaces is truncated at certain high and low frequencies.

We consider normal contact between two linearly elastic
half-spaces ($E_1,\nu_1$ and $E_2,\nu_2$ are Young's moduli and
Poisson's ratios of the solids) possessing rough surfaces
$h_1(x,y),h_2(x,y)$.
Under assumption of frictionless non-adhesive contact, this problem may be replaced by contact
between a rigid surface with an effective roughness $h=h_1-h_2$ and
an elastic flat half-space with effective Young's
modulus~\cite{johnson1987b} 
\be
E^*=E_1E_2/((1-\nu_1^2)E_2+(1-\nu_2^2)E_1).
\label{eq:effective_ym}
\ee
This substitution is common and enables to use numerical methods, which are simpler
than those needed for the original formulation.


In Section~\ref{sec:overview} we give an overview of analytical and numerical models of rough contact.
In Section~\ref{sec:roughness} we discuss the generation of rough surfaces with prescribed properties, also we demonstrate the role of cutoffs 
in surface spectrum on the Gaussianity of resulting roughness. Equations linking Nayak's parameter and asperity density with
the Hurst exponents and cutoff wavenumbers are derived (see also~\ref{app:alpha}).
In Section~\ref{sec:num_model} the numerical model and the set-up are briefly outlined.
The evolution of the real contact area at light loads is analyzed and compared to analytical models in Section~\ref{sec:ca_light}.
General trends in the contact area evolution from zero to full contact are discussed in Section~\ref{sec:ca_light_full}.
Asymptotics of the contact area near the full contact is investigated in Section~\ref{sec:ca_full}.
In Section~\ref{sec:num_error} we propose an estimation of error bounds of the contact area in numerical simulations
and experimental measurements. In Section~\ref{sec:disc} we discuss the obtained results and prospective work.

\section{\label{sec:overview}Overview of mechanical models of rough contact}

\subsection{Analytical models}

Two classes of analytical models exist.  The first class is based on
the notion of asperities (summits of the surface at which $\nabla h =
0$). The pioneering work by Greenwood and Williamson
(GW)~\cite{greenwood1966prcl} was followed by more elaborated models
refining geometrical and statistical aspects of the GW
models~\cite{mccool1986w,bush1975w,greenwood2006w,thomas1999b,carbone2008jmps}.
The statistical properties of asperities (e.g., joint probability
density of heights and curvatures) are often derived from the random
process description of rough surfaces~\cite{nayak1971tasme} or may be measured directly; tips of
asperities may be assumed spherical or elliptical, with constant or
varying curvature. 
Note that the progress in asperity based models is strongly associated with Nayak's extension~\cite{nayak1971tasme} 
of Longuet-Higgins studies~\cite{longuethiggins1957rsla} on statistical properties of random surfaces.
Well thought overviews of asperity based models complemented by new insights may
be found in~\cite{carbone2008jmps,paggi2010w}.  

The main limitation for this class of models is that the considered pressure should be very small.
This limitation arises from the following approximations: (1)
contacting asperities are assumed to not interact with the remaining
surface through elastic deformation of the substrate, i.e.  the change
in vertical position of asperities neighboring to a contacting
asperity is not taken into account; (2) asperities coming in contact
are assumed to have a constant curvature for all considered loads; (3)
consequently, the coalescence between adjacent growing contact zones
is not possible. These limitations are very strong and the extension
of these models beyond infinitesimal contact may be quite inaccurate,
especially when junction between contacts occurs (see discussions in~\cite{nayak1973w,greenwood2007w}). 
The notion of asperity in itself is also criticable, as according to fractal
nature of roughness each ``asperity'' has other ``asperities'' at its
tip and so on~\cite{greenwood2001m}, moreover, the scales of these
``asperities'' are not separable.  
Nevertheless, these models 
survived severe criticism and are still well alive because of their relative simplicity and
computational attractiveness. 
The interaction between asperities may be included by means of semi-analytical modeling~\cite{paggi2010w,yastrebov2011cras}, though the problem of
coalescence cannot be solved in this framework. 
Another model was suggested in~\cite{afferrante2012w}, where the authors take into account the elastic interactions and replace two coalescing asperities
by a single asperity with effective properties. Nevertheless, these approaches remain quite approximate.

The second class of models was initiated by Persson~\cite{persson2001jcp,persson2001prl}.  
These models do not use the notion of asperities and rely on the relation between
surface heights and contact pressure distributions in the limit of full contact.
The author obtains a diffusion equation for the probability density (it acts as density of diffusing material) of the
contact pressure (acts as spatial coordinate) depending on the variance
of the surface roughness\footnote{In Persson's model the variance of
  the surface roughness is connected with the magnification, which is the ratio between the upper and lower cutoffs $k_s/k_l$
  in the surface spectrum.} (acts as time). 
This equation was extended to finite pressures and partial contacts by imposing a boundary
condition~\cite{persson2002prb}: probability density of the zero local contact pressure is zero for all applied pressures.
This model was compared with numerical results: rather successfully in terms of contact
pressure distribution~\cite{campana2007epl,campana2008jpcm,putignano2012ijss,putignano2013ti}
and less successfully in prediction of the contact
area~\cite{hyun2004pre,yang2008jpcm,yastrebov2012pre}.
The validity of this model for partial contacts was
criticized~\cite{manners2006w}; the same authors give a simpler
form for Persson's diffusion equation and discuss possible improvements of the model.

\subsection{\label{ssec:num_methods}Numerical models}

Since last decades several groups carried out numerical simulations of mechanical interaction between rough surfaces.
Most of these studies are limited by elastic frictionless contact for which Johnson's assumption is valid (see Section~\ref{ssec:mechanics}).
This relatively poor problem (which includes only basic mechanical contact
and ignores other relevant surface and bulk phenomena) remains attractive for
research as reliable results for the evolution of the contact area
have not yet been obtained and generalized; another reason is that the results can be compared with predictions of existing analytical models.
Some work was carried out on elasto-plastic contact between rough surfaces~\cite{yan1998jap,pei2005jmps,nelias2005jt,gao2006prc,yastrebov2011cras}
demonstrating, for example, a significantly more linear evolution of the contact area than for elastic surfaces~\cite{pei2005jmps}.
Another interesting observation is that for high pressures at fully plastic deformations the shape of an asperity does not affect the contact pressure evolution
for a single asperity (see \cite{mesarovic1999prs,song2013mm} and \cite[Fig. 19]{yastrebov2011cras}). 
However, one has to remember that at microscopical asperity scales, simple
elastic-plastic models are not valid and the metal microstructure may
affect significantly the contact behavior.  Also, it is important to note that the Johnson's assumption is not valid for nonlinear
materials and rigorously, the obtained results cannot be simply extended for the contact between two deformable elasto-plastic solids.

Elaborated numerical modeling of elastic frictionless contact using the finite element model (FEM) was first carried out on rough surfaces, whose roughness is preserved down to the discretization scale~\cite{hyun2004pre,pei2005jmps}. 
These works reported many results and gave important insights of statistical nature, 
which influenced several followed up studies.
Nevertheless, the numerical model used for these studies includes only one node per asperity, thus the mechanical response of each contacting asperity was
altered (see mesh convergence study, for example, in~\cite[Fig.4-5]{yastrebov2011cras}) and the contact area was overestimated.
This fact prevented the model to draw precise results on the local behavior of separate contact zones, but allowed to obtain approximate distribution of contact
clusters. This drawback was rapidly recognized and corrected in the majority of successive numerical studies by introducing the shortest wavelength 
that are significantly longer than the surface discretization ($\lambda_s \gg \Delta x$, where $\Delta x$ is a mesh step), though still some studies report results~\cite{pastewka2013pre} for $\lambda_s=2\Delta x$\footnote{Often reported values for the high frequency cutoffs $\lambda_s=\Delta x$ or $k_s=L/\Delta x$ are meaningless as the wavelength equal to the spacing between nodes cannot be represented (see e.g.~\cite{prandoni2008b}); it would be more rigorous to put $\lambda_s=2\Delta x$ or $k_s=L/(2\Delta x)$ as we did in Table~\ref{tab:num_studies}.}.

The successive studies used numerically more attractive techniques (similar to a boundary element method) allowing to discretize only the surface of the contacting solid and not the bulk,
namely the Green's Function Molecular Dynamics (GFMD) \cite{campana2006prb}, Boundary Element Method (BEM)~\cite{putignano2012ijss}, and Smart Block Molecular Dynamics (SBMD)\footnote{The SBMD relies on a coarse discretization of the bulk retaining a coherent Molecular Dynamics description of the surface.}~\cite{yang2006epje}.
Thus these methods were able to increase the mesh density to perform more accurate studies.
In Table~\ref{tab:num_studies} we summarize the recent numerical studies, the employed numerical methods, used cutoffs and discretization.
Among the performed simulations, those for which the lower wavenumber cutoff\footnote{Precisely, we mean the wavenumber, from which the power spectral density starts to decrease as power law of wavenumber.} $k_l=1,2$ were carried out on non-Gaussian surfaces with several major asperities;
such surfaces differ significantly from real random rough surfaces. The studies, which were carried out for $L/(\Delta x k_s)<8$ ,
suffer from an imprecise local response, i.e. the asperities are poorely discretized, that also affect the mechanical response ($L$ is the length of a side of the simulated square surface).
In bold we highlighted the pairs $\{k_l,k_s\}$ rendering, we believe, mechanically meaningful results.
Besides the cutoffs, the results reported in~\cite{yang2006epje} (SBMD) and in~\cite{pohrt2012prl} (BEM) suffer from numerical/set-up errors 
and have to be interpreted with prudence; the former model cannot reproduce a simple Hertz contact test, the latter uses a set-up possesing stress singularities
on borders of the simulated contact zone (see also critics in~\cite{pastewka2013pre}).

\begin{table}
\begin{center}
\begin{tabular}{lcccc}
  Article  & Method  & $k_l$ & $ L/(\Delta x k_s)$ & Mesh $L/\Delta x$\\\hline
  $^*$\cite{hyun2004pre} 			& FEM  		 & $1$ & $2$ & $64$-$512$ \\ 
  \cite{pei2005jmps} 				& FEM  		 & $1$ & $2$ & $64$-$512$ \\
  $^*$\cite{hyun2007ti}  			& FEM  		 & $1,\mathbf8$ & $2,\mathbf2^\dag, \mathbf8$ & $512$\\
  $^*$\cite{campana2007epl} 			& GFMD 		 & $2$ & $2^\dag$ & $2048$\\
  \cite{griebel2007b} 				& SBMD 		 & $3$ & $2,7,18$ & $647$\\
  \cite{campana2008jpcm} 			& GFMD 		 & $1$ & $2$-$64$ & $4096$\\
  \cite{campana2008pre}			& GFMD 		 & $1$ & $2$-$64$ & $4096$\\
  \cite{campana2011jpcm} 			& GFMD 		 & $2\ddag$ & $32$   & $2048$\\
  \cite{almqvist2011jmps}			& GFMD,BEM 	 & $2\ddag$ & $64$ & $2048$\\
  \cite{almqvist2011jmps}			& SBMD 		 & $2\ddag$ & $16$ & $512$\\
  \cite{akarapu2011prl} 			& MD,GFMD	 & $1$	& $\approx 5$ & $84$-$340$\\
  \cite{dapp2012prl}				& GFMD 		 & $1$-$\mathbf{4}$ 	& $2$-$\mathbf{8}$	      & $4096$\\
  \cite{putignano2012ijss}[7]$^\star$		& BEM		 & $1$	& $16,32$	& $\le2048$\\
  $^*$\cite{putignano2012jmps} [7]$^\star$ 	& BEM		 & $1$	& $16$-$64$	& $\le2048$\\
  $^*$\cite{yastrebov2012pre} [12]$^\star$	& BEM		 & $1$-$\mathbf{16}$	& $2$-$\mathbf{32}$ & $1024$\\
  $^*$\cite{pohrt2012prl} [60]$^\star$		& BEM 		 & $1$		& $2$	   & $2048$\\
  \cite{pastewka2013pre}			& GFMD   	 & $1$-$8$	& $2$	   & $8912$\\
  \cite{putignano2013ti} [7]$^\star$		& BEM		 & $1$		& $2$-$128$	& $128$-$2048$\\
  $^*$\cite{prodanov2014tl} [4]$^\star$        & GFMD           & \textbf{ext.}$^\wp$   & \textbf{ext.}          & $32768$-$131072$\\
  \hline\footnotesize
\end{tabular}
\end{center}
\caption{\label{tab:num_studies}List of recent numerical studies of the rough contact; relevant parameters $k_l$ and the ratio $L/(\Delta x k_s)$ as well as the discretization $L/\Delta x$ are also given. In bold we highlighted mechanically meaningful combinations of $k_l$ and $k_s$. \newline
\emph{Footnotes:} With $^*$ we mark articles studying the coefficient of proportionality $\kappa$.
$^\dag$Surfaces are smooth, the high value of $k_s$ is due to a bi-quadratic or B\'ezier interpolation between experimental points.\newline
$^\ddag$This value was not mentioned explicitly, we deduced it from the form of the surface and 
its power spectral density, see \cite[Fig.3]{campana2011jpcm} and \cite[Fig.4]{almqvist2011jmps}.\newline
$^\star$Number of statistically equivalent surfaces that was used to average the results; if the number is not listed, the number of realizations is not reported in the paper and we assume that a single surface was used.\newline$^\wp$By ``ext.'' we imply, following the authors~\cite{prodanov2014tl},
the values extrapolated to infinitesimal limit.}
\end{table}

In many listed studies\footnote{Except the following articles~\cite{hyun2007ti,dapp2012prl,pastewka2013pre,yastrebov2011cras,yastrebov2012pre,prodanov2014tl}} the longest wavelength in the surface spectrum was equal to the domain size $\lambda_l=L$ (or simply $k_l=1$) that renders considered surfaces strictly non-Gaussian (see Section~\ref{sec:roughness}).
In such a case, the generated rough surface has only a few large ``macro asperities'', that cluster the contact zones. 
This effect coupled with periodic boundary conditions changes the elastic response of the surface, and consequently the topology and the value of the real contact area. 
The signature of numerical simulations with $k_l=1$ is a pronounced clustering of contact zones close to peaks of ``macro asperities'', see for instance,
\cite[Fig. 12]{hyun2004pre},\cite[Fig. 3]{pei2005jmps}, \cite[Fig. 7]{almqvist2011jmps},
\cite[Fig. 3]{putignano2013ti}, \cite[Fig. 6]{putignano2012ijss}, \cite[Fig. 10-11,13-14]{putignano2012jmps}, \cite[Fig. 7]{almqvist2011jmps}, and Fig.~\ref{fig:area_evolution_pic} in the current article.

\begin{figure}
 \includegraphics[width=1\textwidth]{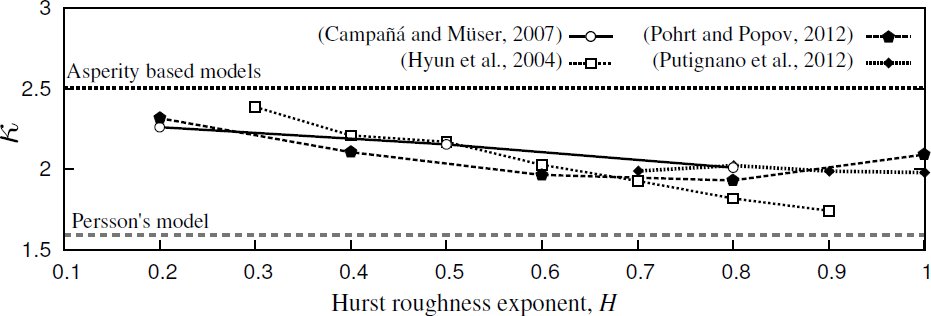}
 \caption{\label{fig:kappa_others}Results of recent numerical studies on the coefficient of proportionality $\kappa$ between the contact area and the pressure.} 
\end{figure}

It is often reported that the contact area increases almost proportionally to the contact pressure at light contact (up to $15\%$ of contact area). Some studies listed in Table~\ref{tab:num_studies} (marked with $^*$) also report results on the estimation of proportionality coefficient between the normalized pressure and the contact area, denoted as $\kappa$.
In Fig.~\ref{fig:kappa_others} we summarize the previous analysis of the $\kappa$ coefficient. 
Some researchers~\cite{hyun2004pre,campana2007epl} found that $\kappa$ decreases with an increasing Hurst exponent,
\cite{pohrt2012prl} observed a non-monotonic behavior of $\kappa$ and \cite{putignano2012jmps} found that there is no pronounced dependence of $\kappa$ on the Hurst exponent.
Recently, we also demonstrated that for accurately resolved mechanics and for Gaussian surfaces, the coefficient of proportionality is independent of the Hurst exponent. 
We consider also that the results depicted in Fig.~\ref{fig:kappa_others} are affected by 
(1) non-Gaussianity of rough surfaces due to $k_l=1$~\cite{hyun2004pre,campana2007epl,putignano2012jmps}, \cite[supplemental material]{pohrt2012prl} and 
(2) inaccuracy in estimating the coefficient of proportionality $\kappa$.
Rigorously speaking, the relation between contact area and pressure may be assumed linear only at unrealistic contact area fractions; at greater fractions this relation is nonlinear~\cite{greenwood2006w,carbone2008jmps,paggi2010w,yastrebov2012pre}. Thus,
the estimated coefficient of proportionality (a tangent or a secant) will strongly depend on the range of contact areas at which it is estimated.
So we suggest that the representation of results in form of the evolution of inverse mean contact pressure $E^*\sqrt{\langle|\nabla h|^2\rangle} A/A_0p_0$ (which tends to $\kappa$ as $A/A_0\to0$) should be preferable to simple contact area evolution with pressure. This representation was adopted in~\cite[Fig. 2,3]{hyun2007ti} and in~\cite[Fig. 3]{yastrebov2012pre}.

Concerning the importance of the lower cutoff, first in~\cite{hyun2007ti} authors made a remark on the significant influence of the lower cutoff wavenumber on the response of an elastic rough surface, and demonstrated an alteration of $\kappa$ for different $k_l$ as well as its nonlinear evolution. 
Comparing figures from~\cite[Fig. 4]{hyun2007ti}, one sees that in Fig.4a-c (plotted for $k_l=1$) a pronounced localization of contact zones appears
whereas the localization is absent in Fig. 4d (plotted for $k_l=8$). 
However, a rigorous explanation of the observed
discrepancies was not given, partly because of strong fluctuations in measured quantities (a single surface realization was used).
Next, in \cite{yastrebov2011cras} authors introduced the notion of a representative surface element by the proximity of its height distribution to a reference Gaussian distribution
obtained for a large surface.
Later, we connected the notion of the representativity with the lower cutoff wavenumber in~\cite{yastrebov2012pre} and results were obtained demonstrating the strong dependence of the proportionality coefficient $\kappa$ and other relevant quantities on the representativity of considered surfaces.
Later, Barber~\cite{barber2013pre} highlighted the importance to have $\lambda_l\ll L$ for finite size systems, which certainly holds for infinite periodic systems.
In~\cite{dapp2012prl} the authors used $2<k_l<8$ to study the leakage and percolation through a contact interface, as it is clear that the topography
of the residual volume (through which the liquid can pass) is even more sensitive to the the lower cutoff wavenumber than the mechanical contact.
However, only recently the importance to have $k_l\gg1$ was fully recognized~\cite{pastewka2013pre}\footnote{Unfortunately, another extreme was reached in this study, the upper cutoff wavenumber $k_s$ was chosen too high $L/(\Delta x k_s)=2$},\cite{prodanov2014tl}.
However, this observation was not connected with Gaussianity of generated surfaces.
It is important to note that averaging of results for many realizations with low $k_l$ does not converge to the results obtained for high $k_l$.
The study of averaged mechanical response of rough surfaces is inherently similar to the homogenization procedure applied to representative volume element~\cite{kanit2003ijss}.
In this light, the lower cutoff $k_l$ may be considered as the size of representative surface element and $k_s/k_l$ as its detailization.
The drawn conclusions are relevant for all random systems with long range interactions (elastic, gravitational, electrostatic).
As stated in~\cite{yastrebov2012pre} and as discussed in the current article, the relevant mechanics and scaling for different cutoff wavelength remain unclear
and require further investigations.

Along with the mean surface gradient, Nayak's parameter $\alpha$ is the main characteristic of rough surfaces~\cite{nayak1971tasme} (see also Section~\ref{sec:roughness}).
It also controls the mechanical response of rough surfaces~\cite{bush1975w,greenwood2006w,carbone2008jmps,paggi2010w}.
Strangely, in spite of its crucial importance, this parameter was not considered in the totality of cited studies\footnote{With the notable exception of~\cite{paggi2010w}, in which a quite different numerical technique was used and thus this work was not listed in Table~\ref{tab:num_studies}.}.
In this paper we aim to introduce an accurate study of Nayak's parameter in numerical mechanics of rough contact.

\section{\label{sec:roughness}Some properties of generated rough surfaces}

\begin{figure}[htb!]
 \begin{center}
  \includegraphics[width=0.8\textwidth]{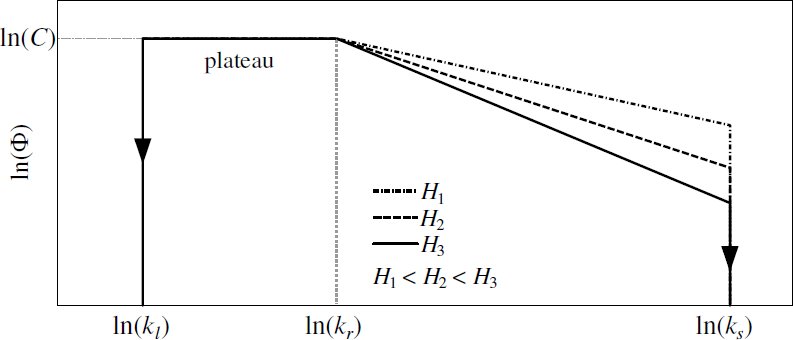}
 \end{center}
 \caption{\label{fig:a:k_lks}Approximated power spectral density of
   rough surfaces with lower $k_l$, roll-off $k_r$ and upper $k_s$ cutoff wavenumbers.}
\end{figure}

For this study we generate artificial rough self-affine surfaces.
This enables us to control all relevant parameters and 
understand how they affect the mechanical behavior of rough surfaces.
To construct \emph{periodic} rough surfaces we use a Fourier based 
filtering algorithm \cite{hu1992ijmtm}. It allows to generate topographies $h(x,y)$
with a power spectral density (PSD) $\Phi(|\mathbf k|)$, which is given approximately by the following equation (see Fig.~\ref{fig:a:k_lks}):
\bes \Phi(|\mathbf k|) = \begin{cases}
   C &,\mbox{ if } k_l < |\mathbf k| < k_r;\\
  C (|\mathbf k|/k_r)^{-2(1+H)} &,\mbox{ if } k_r \le |\mathbf k| \le  k_s;\\ 
  0 &,\mbox{ otherwise}.
            \end{cases}
\ees 
where $C$ is the constant determining the roughness amplitude and $\mathbf k$ is the wavevector. 
Thus five parameters can characterize the produced surfaces: $k_l,k_r$ and $k_s$
the lower, roll-off and upper cutoff wavenumbers respectively 
(or equivalently the longest, roll-off and shortest
wavelengths $\lambda_l=L/k_l,\lambda_r=L/k_r, \lambda_s=L/k_s$), the root mean
square (rms) of surface gradients $\sqrt{\langle|\nabla h|^2\rangle}$, and finally the Hurst roughness exponent $H$.
In all our simulations the plateau is not present in the spectrum, so $k_r=k_l$.

Nayak's parameter $\alpha$, which characterizes the breadth of the surface spectrum, was introduced in~\cite{nayak1971tasme} through spectral moments: 
\be
\alpha=m_0m_4/m_2^2,
   \label{eq:alpha_mom}
\ee 
where for isotropic surfaces $m_0=m_{00}$, $m_2 = m_{20} = m_{02}$, $m_4 = 3m_{22} = m_{40} = m_{04}$ and 
$$
  m_{pq} = \iint\limits_{-\infty}^{\quad\infty} k_x^pk_y^q \Phi^s(k_x,k_y)\,dk_xdk_y.
$$ 
For an isotropic surface, $\alpha$ can be expressed through the ratio $\zeta=k_s/k_l$ (referred as magnification in Persson's model)
and the Hurst exponent as (see~\ref{app:alpha})\footnote{In~\ref{app:alpha} we derive an equation linking Nayak's parameter, Hurst exponent and cutoff wavenumbers for surfaces with and without plateau in the PSD. We derive also a formula for the mean density of asperities.}:
\be \alpha(H,\zeta) =\frac{3}{2}
\frac{(1-H)^2}{H(H-2)}\frac{(\zeta^{-2H}-1)(\zeta^{4-2H}-1)}{(\zeta^{2-2H}-1)^2}.
\label{eq:alpha} 
\ee 
Note that the Nayak's parameter does not depend on the width of the bandwidth $k_s-k_l$ but on the ratio $k_s/k_l$.
For a high enough ratio $\zeta$, a simple asymptotic may be used: 
$$\alpha\sim\zeta^{2H}.$$ 
Consequently, for a surface with an infinite breadth spectrum ($\zeta\to\infty$) the Nayak's parameter also tends to infinity.
Even though the Hurst exponent is simpler to determine experimentally for realistic surfaces, we 	
conjecture that it is possible that the mechanical response of surfaces may depend only on the Nayak's parameter rather than Hurst exponent.

Another characteristic, which enters asperity based model, is the density of asperities $D$; it can be also expressed through the cutoff wavenumbers and the Hurst exponent (see~\ref{app:alpha}):
\be 
   D=\frac{\sqrt3}{24\pi}\frac{1-H}{2-H}\frac{\zeta^{4-2H}-1}{\zeta^{2-2H}-1}k_l^2.
   \label{eq:D}
\ee
For a high enough ratio $\zeta$, a simple asymptotic may be used $D \sim k_s^2$.

The computation of the root mean square gradient is crucial for comparing numerical results with analytic theories.
An accurate computation of this quantity may be done only in the Fourier space, giving the following expression
$$
 \sqrt{\langle |\nabla h|^2\rangle} = \sqrt{m_{02}+m_{20}},
$$
which is valid both for isotropic and anisotropic surfaces.
The value computed by finite differences~\cite{paggi2010w} depends on the discretization of the surface and
thus may underestimate significantly the real value; it is especially important for surfaces in which the shortest wavelength is close to
the discretization step~\cite{hyun2004pre,pei2005jmps}.

\begin{figure}[ht!]
 \includegraphics[width=1\textwidth]{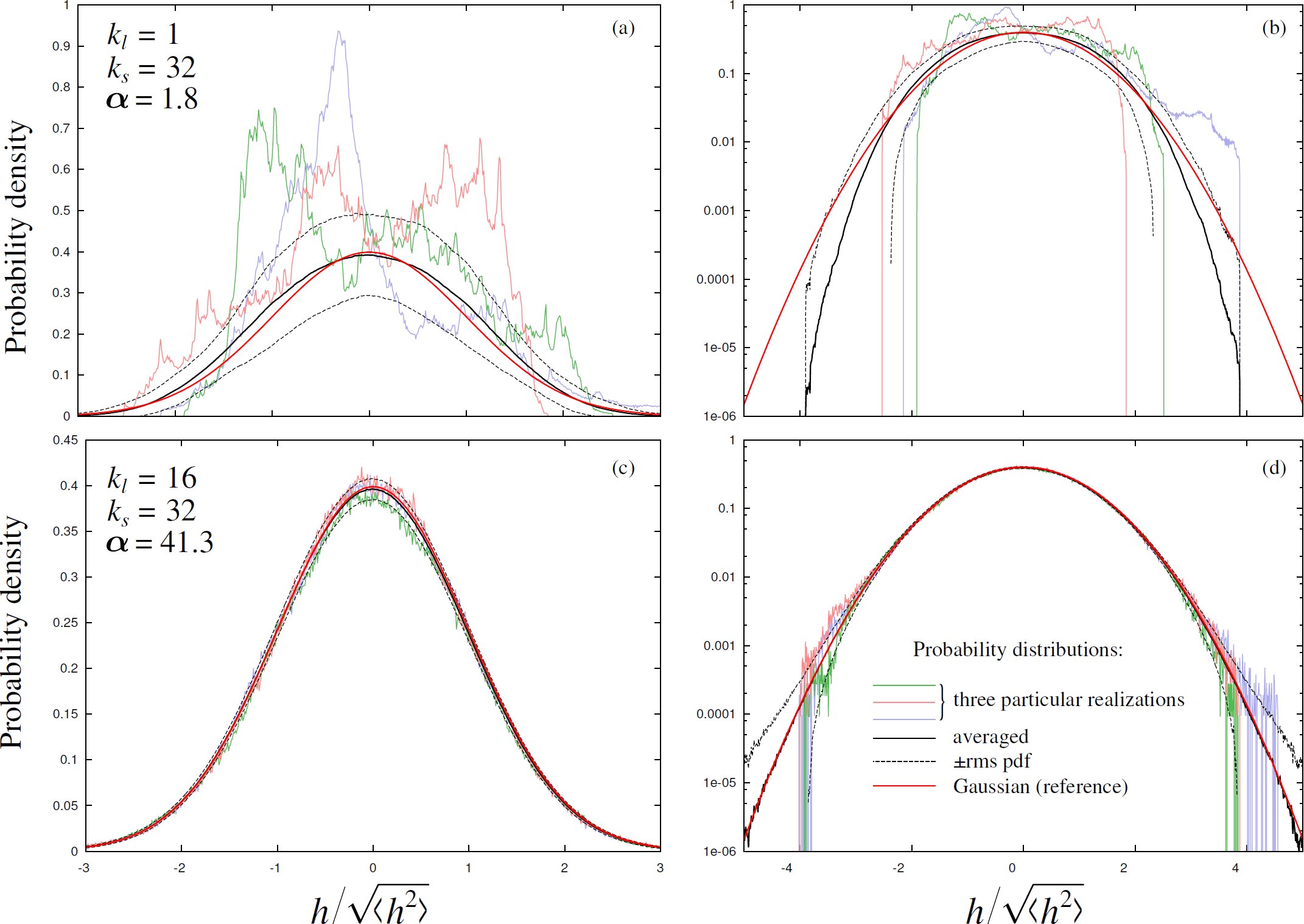}
 \caption{\label{fig:normality}Surface heights distribution for
   different cutoff wavenumbers (a,b) $k_l=1, k_s=32$ (c,d) $k_l=16, k_s=32$. Distributions are depicted for
   for three particular surfaces (colored oscillating lines), for a distribution averaged over 1000 statistically equivalent surfaces $H=0.8$ (solid black line),
   its standard deviation (dashed black line) and a reference Gaussian distribution (solid red curve); (a,c) linear and (b,d) semi-logarithmic plots.}
\end{figure}

From analytic results~\cite{nayak1971tasme,persson2005jpcm}, it is known that a Gaussianity of heights of rough surfaces is independent of the spectral content.  
It was however demonstrated~\cite{yastrebov2012pre} that for low $k_l$ one cannot approach a Gaussian surface heights distribution for generated surfaces, 
even for very high upper cutoff $k_s\gg1$. 
This inconsistency between numerically generated surfaces and analytic predictions comes from the discreteness of the wavevector's space in
finite systems, i.e. $k_x$ and $k_y$ may take only integer values to ensure the periodicity of generated surfaces.
In Fig.~\ref{fig:normality} we show some surface height distributions averaged over $1000$ surfaces discretized in $1024\times1024$ points.
To make the averaging consistent, each surface height was normalized by the root mean square of heights $\sqrt{\langle h^2\rangle}$; the limits of distribution are fixed
$h\in[-5\sqrt{\langle h^2\rangle}:5\sqrt{\langle h^2\rangle}]$, this interval is divided into $1000$ bins on which the distribution is evaluated. 
The averaged distributions are obtained for different cutoff wavenumbers. 
For $k_l=1$, even though the averaged distribution resembles a Gaussian one, each particular distribution is strictly non-Gaussian (see colored curves in Fig.~\ref{fig:normality}(a)); increasing $k_s$ does not solve the problem; it practically has no effect on the distribution (see additional plots in~\ref{app:height_distr}).
This strong deviation from the normality is revealed through plotting the standard deviation of the distribution function (see black dashed lines in Fig.~\ref{fig:normality}); it is large for $k_l=1$ and decreases rapidly when $k_l$ is increased.
Note that a normally distributed amplitude for each wavevector~\cite{pastewka2013pre} cannot ensure the normality of the resulting surface.
When we increase the lower cutoff to $k_l=4,\; 16$, an accurate Gaussian distribution is obtained 
even for individual surfaces.
It is clear that the \emph{averaged mechanical response of non-Gaussian surfaces} is not equivalent to the \emph{mechanical response of the averaged surface}, whose distribution is Gaussian.
Another important issue is the isotropy of the considered surface, which cannot be ensured if $k_l=1$.
For these reasons, to obtain a meaningful mechanical response of a rough surface, it is crucial to introduce a sufficiently high lower cutoff in the surface spectrum.

\begin{figure}[ht!]
  \includegraphics[width=1\textwidth]{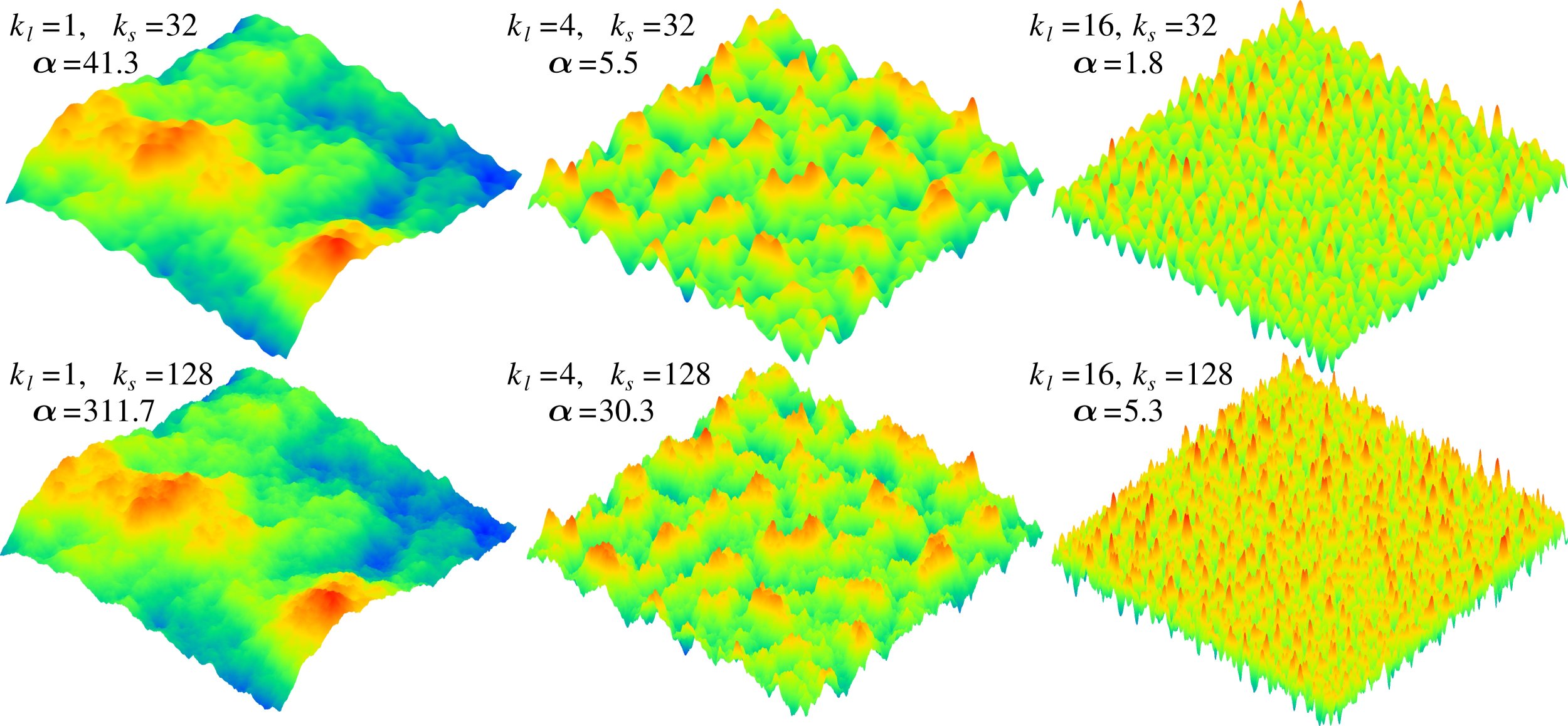}
 \caption{\label{fig:surfaces}Examples of generated rough surfaces for
   $H=0.8$ and different cutoff wavenumbers.}
\end{figure}

In this study, we use nine types of periodic surfaces combining cutoffs $k_l=\{1,4,16\}$
with $k_s=\{32,64,128\}$ for $H=0.8$ with fixed rms gradient $\sqrt{\langle|\nabla h|^2\rangle}=.02$ and fixed size $L=1$ 
(examples of generated surfaces are given in Fig.~\ref{fig:surfaces}).
We analyze three cutoffs at each side to deduce the corresponding trends in mechanical behavior.
The surfaces are discretized in $1024\times1024$ elements.
To obtain statistically meaningful results for each combination of parameters, 50 surfaces are generated and used for mechanical simulations.
The effective Young's modulus~\eqref{eq:effective_ym} is set to $E^*=1$.

\section{\label{sec:num_model}Numerical model}

Since we consider a frictionless non-adhesive contact between linearly elastic half-spaces, the problem may be reduced
to the contact between an effective elastic flat half-space and a rigid rough surface (see Section~\ref{ssec:num_methods}).
To resolve the mechanical boundary value problem with contact constraints, we use the FFT
based boundary element method suggested in~\cite{stanley1997}\footnote{The original formulation of the FFT algorithm~\cite{stanley1997} contains some errors.
The most relevant is that the solution is shifted in Fourier space by one wavenumber.
For the current and previous~\cite{yastrebov2012pre} studies we use a corrected version of this method, which was validated on many cases.}.
The main advantage of the class of boundary element methods compared to finite element methods is its computational attractiveness.
Another advantage is that the periodicity of the solution is inherent, as the solution is defined in Fourier space.
It is also important that the deformed solid is assumed semi-infinite, which is normally not the case in finite element models.

An external pressure $p_0$ is applied in approximately $85-90$ increments up to about $p_0=1.7 E^*$ which corresponds to full contact\footnote{The number of increments and the maximal pressure may differ from one surface to another as the simulation is stopped when all points come in contact.}.
To increase the precision in the evolution of the contact area for small contact areas, the first $50$ loading steps are confined in $p_0\in[0;\,0.06E^*]$, which
corresponds to the contact area fraction $A/A_0 \in [0;\, \approx14\%]$.

\section{\label{sec:ca_light}Contact area at light pressure}

\subsection{\label{ssec:light_analytics}Light contact: analytical models}

For infinitesimal pressure, analytical models predict a linear growth
of the real contact area with external pressure 
\be\label{eq:linear_a}
\frac{A}{A_0} = \frac{\kappa}{\sqrt{\langle|\nabla h|^2\rangle}}
\frac{p_0}{E^*}. 
\ee 
The coefficient of proportionality was found to be
$\kappa=\sqrt{2\pi}$ in all elaborated asperity based
models~\cite{bush1975w,greenwood2006w}, for a detailed review see~\cite{carbone2008jmps}.  
Note, that this equation holds only for very small contact area fractions, which
corresponds to a separation between mean planes that tends to infinity.
As the convergence to this asymptotic behavior is very slow, it cannot be verified neither numerically nor experimentally.
For example, as noticed in~\cite{yastrebov2012pre} for a simplified elliptic model~\cite{greenwood2006w}, for a surface with Nayak's
parameter $\alpha=10$ and area fraction $A/A_0\approx10^{-5}$, the
coefficient $\kappa$ computed from Eq.~\eqref{eq:linear_a}
underestimates the asymptotic value by $12\%$; if the real contact
area is smaller by five orders of magnitude $A/A_0\approx10^{-10}$, the
computed $\kappa$ underestimates the asymptotic value by $4\%$. 
The error becomes significantly higher for higher Nayak's parameters.
For more realistic separations/pressures, asperity based
models~\cite{greenwood2006w,carbone2008jmps} predict that the
evolution of the contact area is strongly dependent on Nayak's
parameter: the higher it is, the smaller the contact area for a given
pressure and hence the more it deviates from the asymptotic prediction. 
We recall that the two main assumptions of asperity based models are that
(i) the elastic deformation of the substrate is not taken into
account, i.e. a contacting asperity is assumed not to affect the
neighboring asperities, (ii) asperity's curvature is constant, so the contact zones associated with
different asperities never merge. Both assumptions are justified for
infinitesimal contact, for which the model should be accurate.

We recall that Persson's model~\cite{persson2001jcp,persson2001prl}, on the contrary, does not use the notion of asperities.
This model is based on the evolution of the probability density of the contact pressure with increasing magnification $\zeta=k_s/k_l$.  
Rigorously speaking, Persson's model does not depend on $\zeta$ but only on the variance of the
contact pressure $V(p)$~\cite[see Eqs.~(8),(22)]{manners2006w}, that is
proportional to the mean squared slope at full contact:
$$ V(p) = \frac14E^{*2}\langle |\nabla h|^2\rangle.$$ 
So the rms gradient is the only characteristic of rough surfaces that
is used by Persson's model, in contrast to asperity based models, which 
require the spectrum breadth given by Nayak's parameter $\alpha$ to provide predictions at realistic pressures.  
For the contact between nominally flat and really rough surfaces, Persson's model gives the
following form for the evolution of the real contact area
\be\label{eq:area_persson} 
\frac{A}{A_0} =
\mbox{erf}\left(\frac{p_0}{E^*}\sqrt{\frac{2}{\langle|\nabla
    h|^2\rangle}}\right).
\ee
For infinitesimal pressure it can be replaced by its asymptotic form 
\be\label{eq:area_persson_as}
\frac{A}{A_0} = \frac{\sqrt{8/\pi}}{\sqrt{\langle|\nabla h|^2\rangle}}
\frac{p_0}{E^*}, 
\ee
since $\lim\limits_{x\to0}\mbox{erf}(x)/x\xrightarrow[x\to0]{} 2/\sqrt{\pi}$. 
Note that the only difference
between asymptotic predictions of asperity based models and Persson's model is the coefficient of proportionality $\kappa$.  In
Persson's model $\kappa=\sqrt{8/\pi}\approx 1.6$, which is $\pi/2$
lower than in asperity based models.  For contact areas up to $10-15\%$, 
the Persson's prediction~\eqref{eq:area_persson_as} only slightly deviates from its asymptotics and can be with a good confidence considered as linear.
The main assumption of Persson's model is that it is derived for
the case of full contact and then extended to partial contact, which
is not fully justified~\cite{manners2006w}. Yet this model is precise at full contact.
We recall that the Nayak's parameter $\alpha$ enters in asperity based models, because the mean asperity
curvature increases with the height of the asperity (more precisely, with the vertical coordinate of the asperity's summit, see~\cite[Fig. 5]{nayak1971tasme}).
The smaller $\alpha$, the stronger the difference between the mean curvature and the curvature of highest asperities, 
which are the only asperities that come in contact at light pressures.
In the limit of $\alpha=\infty$ the mean curvature of asperities for a given range of heights does not depend any longer on the height~\cite[Eq. (65)]{nayak1971tasme}.
The fact that this evolution in the mean asperity curvatures with heights is not taken into account in Persson's model (when it is extended to partial contact), 
represents simply an additional assumption of this model.

\subsection{\label{ssec:num}Light contact: numerical results}

\begin{figure}[ht!]
 \includegraphics[width=1\textwidth]{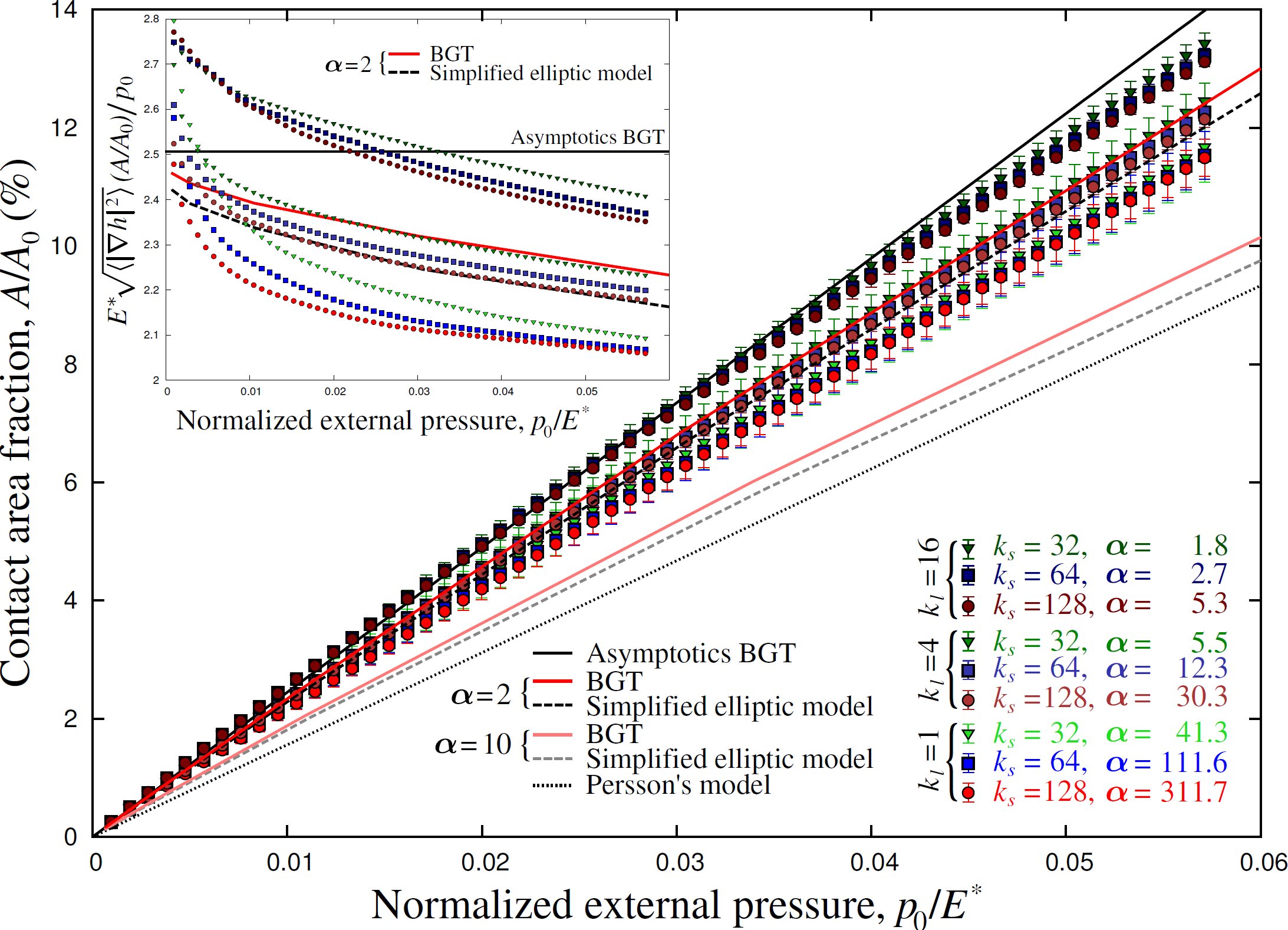}
 \caption{\label{fig:area_num_1}Evolution of the contact area fraction
   with normalized external pressure for different cutoffs (colored points with error bars)
   are compared with Persson's model~\cite{persson2001jcp} and
   asperity based models: BGT~\cite{bush1975w} and simplified elliptic model~\cite{greenwood2006w} for $\alpha=2$ and $\alpha=10$.
   The results for these models are taken from~\cite{carbone2008jmps}. 
   In the inset the secant for the area curve is plotted.}
\end{figure}

\begin{figure}[htb!]\begin{center}
 \includegraphics[width=1\textwidth]{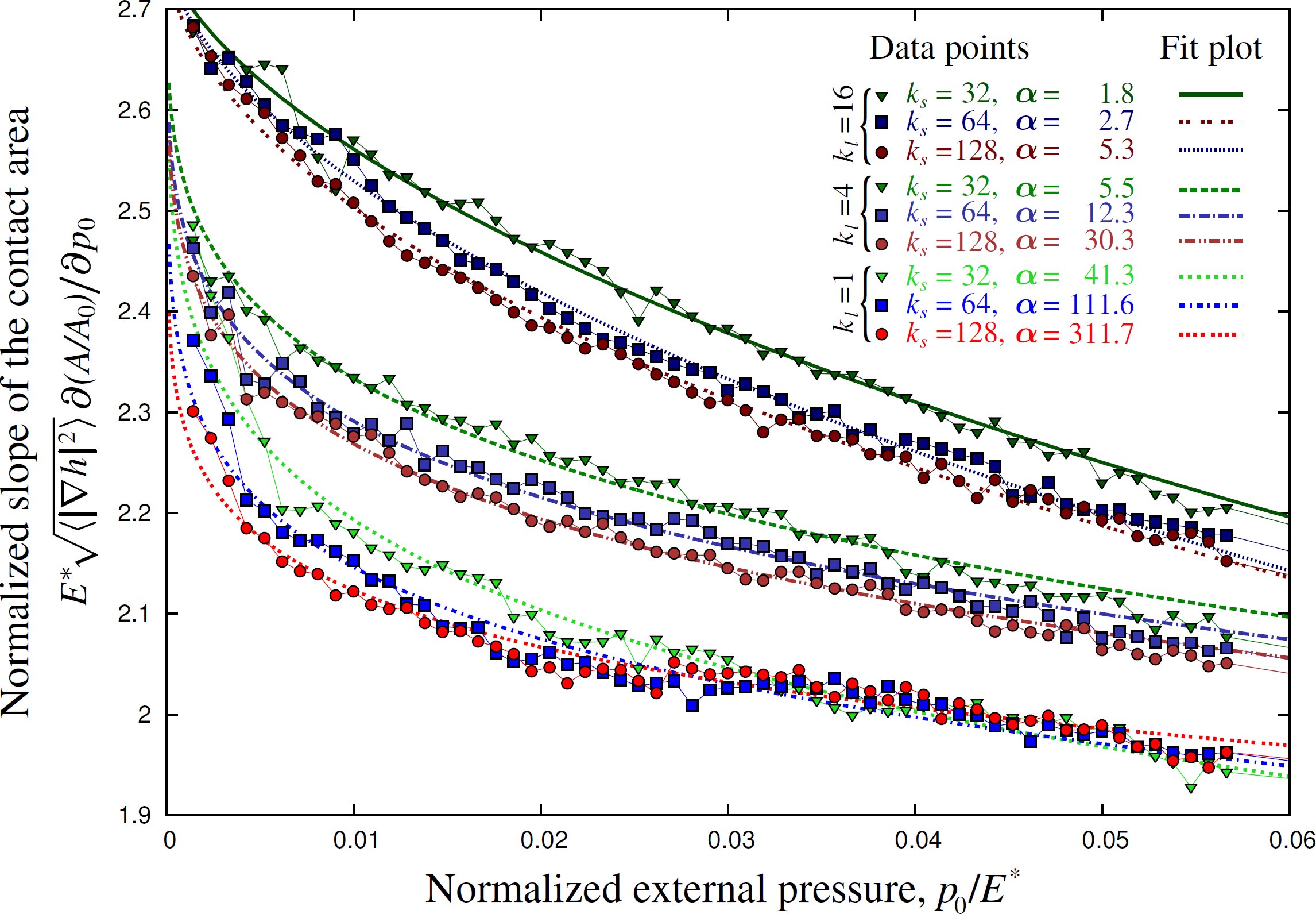}\end{center}
 \caption{\label{fig:area_slope_num}Computed derivative of the
   normalized contact area with respect to the applied pressure (color
   points); fit curves (thick lines); we used Eq.~\eqref{eq:cel_derivative} with
   $\kappa=1.1\sqrt{2\pi}$ and $\beta,\mu$, that are fitted by least square fit.}
\end{figure}

The results for the evolution of the real contact area are depicted in Fig.~\ref{fig:area_num_1} and compared with analytical models.
We recall that every point represents an average over results obtained for 50 different statistically equivalent surfaces subjected to the same load,
over which the error bars (standard deviation) are also computed. 
The higher is the lower cutoff wavenumber $k_l$ (or the shorter is the longest wavelength in the surface spectrum), the higher the predicted area is.
The results for $k_l=16$ are close to the asymptotic prediction of asperity
based models referenced as ``asymptotics BGT''.  
All obtained results are significantly higher than the prediction of Persson's model.
The curves form three groups with respect to the lower cutoff wavenumber $k_l=16, 4, 1$ (see also Fig.~\ref{fig:area_slope_num} and the inset in Fig.~\ref{fig:area_num_1}).
Within each group, curves with greater Nayak's parameter $\alpha$ predict smaller contact areas.
However, one should take into account that for higher $k_s$, the discretization of the shortest wavelength
is coarser, i.e. the number of nodes per shortest wavelength is simply $N_s = N/k_s$, where $N$ is a number of nodes per side $N=L/\Delta x$. 
So we do not exclude that this dependence may be slightly altered by numerical errors. 
Nevertheless, for cutoff wavenumbers $k_s=32$ and $k_s=64$ (which correspond to $N/k_s=32$ and $N/k_s=16$, that are considered sufficient to provide accurate results~\cite{putignano2013ti}), the same trend is observed.
Also, note the considerable difference between the results of the asperity based models predicted for $\alpha=2$ and $\alpha=10$ (BGT~\cite{bush1975w} and the simplified elliptic model~\cite{greenwood2006w}): the contact area for $\alpha=10$ is $\approx30\%$ lower than for $\alpha=2$.
Our results suggest that the effect of $\alpha$ on the contact area is not as strong as predicted by these models.
Compare, for example, curves $k_l=4,k_s=32,\alpha\approx5.5$ and $k_l=4,k_s=128,\alpha\approx30.3$, the $\alpha$ is increased by a factor six, yet the two curves
remain quite close.
Note also that the
standard deviation in the results is higher for smaller $k_s$ and $k_l$: the smallest variance is found for $k_l=16, k_s=128$, the highest is found for $k_l=1, k_s=32$.
To see better the difference between the curves, we plot the secant\footnote{Secant is the slope of the line, that connects the point on the area-pressure curve and the origin.} 
of the contact area curve (Fig.~\ref{fig:area_num_1}, inset). This figure, however, does not enable to determine whether the curves become linear in a certain region or not.

\subsubsection{Coefficient of proportionality $\kappa$}

To assess the difference between numerical results obtained for different cutoffs, we compute the slope of the contact area fraction (Fig.~\ref{fig:area_slope_num}).
Each set of data is fitted by the contact evolution formula suggested in~\cite{yastrebov2012pre}:
\be
\label{eq:cel} \frac{A}{A_0} = \frac{1}{\left[\beta +
    \left(\frac{\sqrt{\langle|\nabla h|^2\rangle} E^*}{\kappa
      p_0}\right)^\mu\right]^{1/\mu}},
\ee 
where $\kappa>0,\beta>0,0<\mu<1$ are the three parameters.
Note that for $p_0/E^* \to 0$ this expression tends to the linear relation~\eqref{eq:linear_a}, for which parameters $\mu$ and $\beta$
disappear, so the parameter $\kappa$ retains its meaning of the coefficient of proportionality between area and pressure at infinitesimal loads;
whereas $\beta$ and $\mu$ depend on geometrical properties of the surface. The complete derivation of Eq.~\eqref{eq:cel} is given in~\ref{app:cel}.
Derivative of \eqref{eq:cel} with respect to $p_0$ gives 
\be
\label{eq:cel_derivative} \ddp{A/A_0}{p_0} =
\frac{\left(\frac{\sqrt{\langle|\nabla h|^2\rangle} E^*}{\kappa
    p_0}\right)^\mu}{ \left(\beta+\left[\frac{\sqrt{\langle|\nabla
        h|^2\rangle} E^*}{\kappa p_0}\right]^\mu\right)^{1+1/\mu} p_0}.
\ee
We used this equation to fit the data points. For that, we chose the minimal value of $\kappa$ that
ensures the convexity of all fitted curves and fitted $\beta$ and $\mu$ by least square error method.
In accordance with predictions of asperity based models, the area slope changes rapidly in the first stages of contact area evolution.
At the same time, it is hard to estimate the contact area at light pressure, as only a few points are in contact.
So it appears, that it is impossible to find a proper value of $\kappa$ for the obtained curves.
Nevertheless, we could suggest a lower boundary for this parameter for which the numerical data points can be fitted appropriately\footnote{By an appropriate fit we imply, that at least the fitted curve (found by least square error method in the region $p_0\in[0;0.06E^*]$) is convex.}: $\kappa \gtrapprox 1.055\sqrt{2\pi}$.
It is however important that we bear in mind that this estimation is quite subjective and is based on the phenomenological equation~\eqref{eq:cel}.
Also the minimal value of $\kappa$ is strongly dependent on the lower cutoff $k_l$.
We will see that this lower boundary can be significantly decreased, if one takes into account the error estimation based on
the perimeter calculation (see Section~\ref{sec:num_error}).

\subsubsection{\label{ssec:contact_evolution}General trends in the contact area evolution}

Besides the debatable $\kappa$ parameter, more general properties can be deduced.
In all obtained curves, the slope of the contact area is a decreasing function of the applied pressure,
and all curves are \emph{convex} in the considered range (up to 10-14\% of the contact area fraction).  
Note that in asperity based models the slope of the area is also convex (can be deduced from data presented in~\cite{bush1975w,carbone2008jmps}), 
whereas in Persson's model the slope decreases as a concave function (see inset in Fig.~\ref{fig:area_evolution_full}).  
Greater $k_l$ or, identically, higher Gaussianity of the surface results in a steeper slope.
At the same time, a greater $\alpha$ or, identically, a greater $\zeta=k_s/k_l$ results in a slighter slope for a given $k_l$.  
However, for different $k_l$ and similar $\zeta=k_s/k_l$, see for example, $\zeta=8$ ($k_l=16, k_s=128, \alpha\approx5.3$ and $k_l=4, k_s=32, \alpha\approx5.5$), 
the area evolutions are quite distinct. 
Hence, for non-Gaussian  surfaces, the Nayak's parameter $\alpha$ cannot solely determine the evolution of the contact area.  
It is also important to note that we do not obtain any convergence in the area evolution curve with an
increasing $k_l$ (approaching a Gaussian surface), (see Fig.~\ref{fig:area_num_1} and \ref{fig:area_slope_num}). 
This trend was observed in~\cite{yastrebov2012pre}, but was not uttered as the Nayak's parameter was not evaluated for the reported calculations.
This non trivial behavior is the subject of the following paragraph.

\subsubsection{\label{ssec:interaction_between_asp}Elastic interactions between asperities}

In asperity based models the interaction between
asperities is neglected, i.e. a contacting asperity deforms on its own, but does
not deform the substrate, thus it does not affect the positions of neighboring asperities. 
This approximation, totally reasonable for infinitesimal contact, strongly alters the results for realistic contact area fractions~\cite[Fig. 15]{yastrebov2011cras}.
Recall that the displacement of points on the surface of a half-space due to a point load
is proportional to the force and inversely proportional to the distance (see e.g., \cite{johnson1987b}).
Without interaction, for a given separation, the force appears to be overestimated compared to the case of interacting asperities.
The trend is inverse for the contact area, that is smaller for non-interacting asperities.
We believe that this interaction between deformable asperities may change the area--pressure curves for different $k_l$.
As in all our simulations the rms gradient is the same, for a given $\zeta$ the density of asperities increases as $k_l^2$ (see Eq.~\eqref{eq:D}), which explains the increasing role of the elastic interactions. Moreover, a low lower cutoff $k_l=1,2$ implies a strong localization of contact clusters, that being repeated by periodic boundary conditions, leads to a specific mechanical response of the contact surface, that is quite distinct from a random one.
Also one may conjecture that the coefficient of proportionality $\kappa$ may be non-unique being affected by different elastic interactions for various surfaces, 
which are not taken into account is asperity based models. Discussion of numerical errors and their correlation with cutoff wavenumbers is given in Section~\ref{sec:num_error}.

\section{\label{sec:ca_light_full}Contact area: from infinitesimal to full contact}

\begin{figure}[t!]
  \includegraphics[width=1\textwidth]{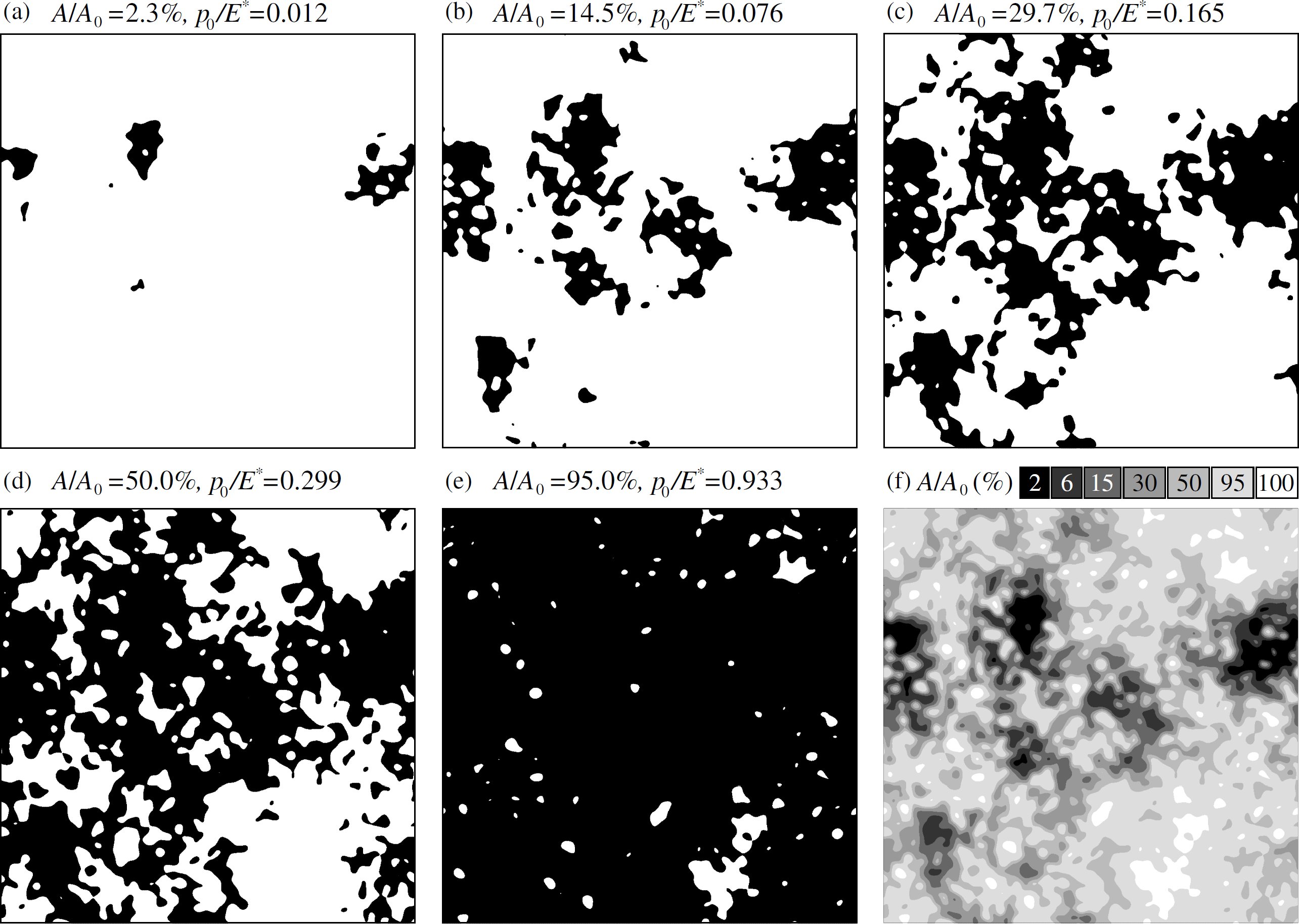}
  \caption{\label{fig:area_evolution_pic}Real contact areas for
    $k_l=1,k_s=32$ at different pressures (a)-(e) and an assembled contour plot (f). 
    Corresponding points are marked at area-pressure curves in Fig.~\ref{fig:area_evolution_full}.}
\end{figure}

\begin{figure}[t!]
  \includegraphics[width=1\textwidth]{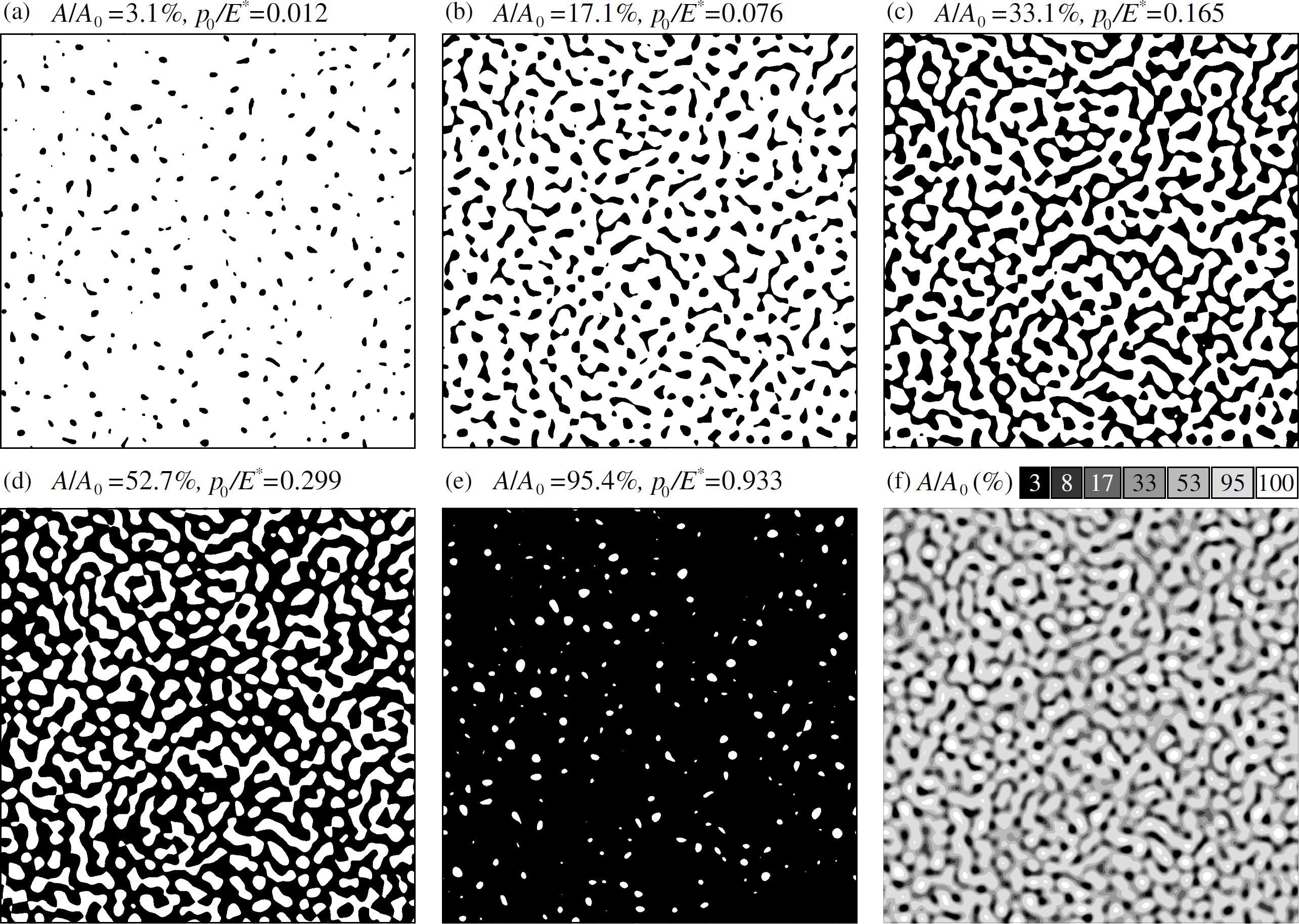}
  \caption{\label{fig:area_evolution_pic2}Real contact areas for
    $k_l=16,k_s=32$ at different pressures (a)-(e) and assembled contour plot (f). 
    Corresponding points are marked at area-pressure curves in Fig.~\ref{fig:area_evolution_full}.}
\end{figure}

Approximations inherent to asperity based models make them not applicable to relatively high contact areas (intermediate contact).
Indeed, at high pressures, the coalescence between contact zones associated with different asperities and their elastic interaction can no longer be neglected.
In our simulations we observed that if the Nayak's parameter is high enough, this coalescence happens even for tiny contact area fractions.
We depict several steps of the contact area topography for surfaces with cutoffs (1) $k_l=1,k_s=32$
(Fig.~\ref{fig:area_evolution_pic}) and (2) $k_l=16,k_s=32$ (Fig.~\ref{fig:area_evolution_pic2}).  
In both cases the anisotropy of local contact zones associated with asperities is visible~\cite{bush1975w,greenwood2006w} as well as the effect of a
strong interaction between contacting asperities, which is more
present in case (1) for which Nayak's parameter $\alpha\approx41.3$ compared to $\alpha\approx1.8$ for case (2) (see, for example, Table~\ref{tab:a:1}). 
Let us remark also that the contact zones are quite evenly distributed in case (2), and clustered in case (1), which is a reliable indicator of a non-Gaussian surface used in (1). 
This clustering accompanied by periodic boundary conditions and long range elastic interactions alters all the results in comparison to truly random surfaces. 
Thus we repeat that a non-Gaussian surface (due to small lower cutoff wavenumber $k_l$) cannot be a good representation of a random rough surface.
Note also that the local contact areas, which are convex at light pressures, rapidly loose this convexity when the pressure is increased;
at greater $\alpha$ they loose the convexity more rapidly.
Increasing the contact area results in a formation of a complex contact topography for which a percolation and leakage problems can be studied~\cite{dapp2012prl}.
But one has to bear in mind that a representative surface element (RSE) for a leakage study is not necessarily equivalent to the RSE used for the elastic problem~\cite{durand2012phd}.


We observe that at light contact area, as classically predicted, the contact area increases due to an increasing (1) number of contact zones and (2) their size.
For higher pressure, the increase in the contact area is associated mostly with the expansion of the existing contact areas. 
This trend is, however, weaker for surfaces with higher Nayak's parameter $\alpha$. 
For small $\alpha$, the growth of the contact area resembles a growth of oil droplets on a flat surface, they expand and form junctions. 
Though, to make this rough comparison slightly more realistic, this flat surface should be imagined as heterogeneous and anisotropic in the sense of wettability.

\begin{figure}[htb!]
  \includegraphics[width=1\textwidth]{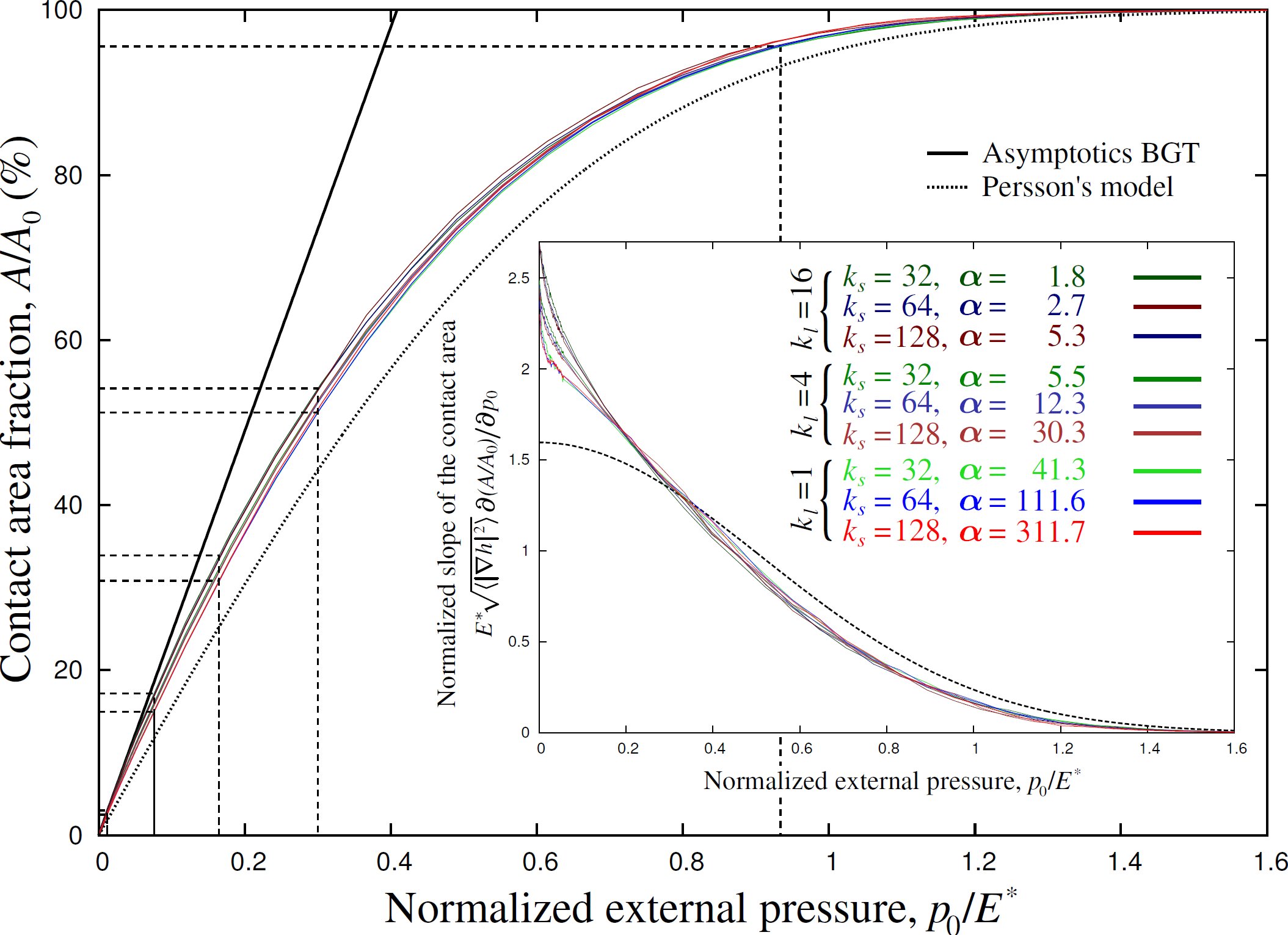}
  \caption{\label{fig:area_evolution_full}Numerical results (color
    solid curves) for the evolution of the contact area fraction from
    zero to full contact compared with the prediction of Persson's
    model (dashed curve) Eq.~\eqref{eq:area_persson}. Asymptotics
    of the BGT model is given as an indication (black solid line).
 We also marked the
    pressures and corresponding areas at which the area topographies are
    depicted in Fig.~\ref{fig:area_evolution_pic},\ref{fig:area_evolution_pic2}.
 In the inset: the
    normalized slope of the contact area as a function of the contact
    pressure is compared with Persson's model.}
\end{figure}

Increase in the contact area with pressure from zero to full contact is depicted in Fig.~\ref{fig:area_evolution_full}. 
All curves are rather close to one another, the influence of cutoffs at this global view seems less pronounced than in the refined study of light contact done in Section~\ref{ssec:num}.  
Nine curves, which we obtained for different cutoffs, form three groups with respect to the lower cutoff wavenumber $k_l$. 
These three groups are easily distinguished up to about $80\%$ of the contact area.

According to Persson's model the contact area evolves as~\eqref{eq:area_persson}. 
This prediction is in a good \emph{qualitative} agreement with the numerical results for the entire range of contact pressures. 
The change in the slope of the contact area is depicted in the inset of Fig.~\ref{fig:area_evolution_full} and compared with Persson's result.
Up to about $50\%$ of contacting area, the numerically predicted slope is steeper than this predicted by the Persson's model and vice versa for greater contact area fractions.
Note also that between surfaces with different cutoffs, the difference in the slope is visible only for light pressures.
For higher pressures, the slope changes approximately in the same manner for all considered surfaces.
The curves with higher $k_l$ seem to yield more consistent curves, which are everywhere convex.

\section{\label{sec:ca_full}Limit of the full contact}

\begin{figure}[htb!]
 \includegraphics[width=1\textwidth]{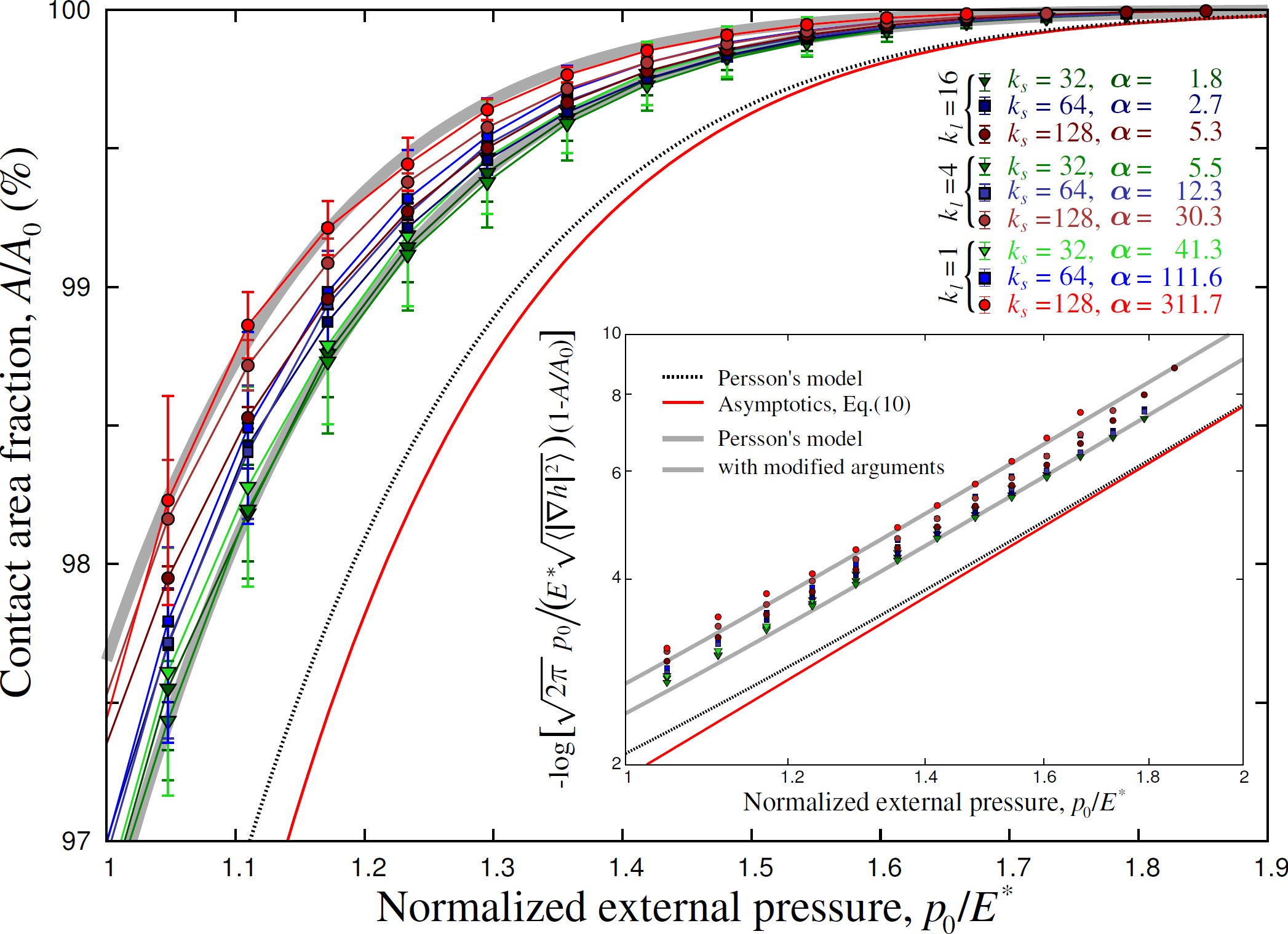}
 \caption{\label{fig:near_full_contact}Contact area approaching full
   contact, comparison between Persson's model (dashed black line),
   Persson's model asymptotics Eq.~\eqref{eq:full_area_approx}
   ($\eta=2$) (red solid line) and numerical results (color points
   with error bars). The Persson's Eq.~\ref{eq:area_persson} fits perfectly the data points if the argument is slightly scaled 
$A/A_0=\mbox{erf}\left(\frac{\beta p_0}{E^*}\sqrt{\frac{2}{\langle|\nabla h|^2\rangle}}\right)$, we plotted this equation for $\beta=1.09$ and $1.16$ (gray thick curves).
   In the inset we plot in logarithmic scale Eq.~\eqref{eq:scaled_area} to demonstrate the agreement between numerical results and Persson's model in terms of the exponent $\eta=2$ in Eq.~\ref{eq:area_persson}.}
\end{figure}

We recall that Persson's model predicts the evolution of the contact area according to Eq.~\eqref{eq:area_persson}.
It is important to note that although Persson's model is developed first for the full contact, 
the full contact is never reached for a finite pressure regardless the breadth of the surface spectrum\footnote{
Although it may seem strange, it does not contradict continuum mechanics, 
as far as the number of modes present in the surface spectrum is infinite. 
This infinite number of modes, never possible in numerical models, ensures the normality of the surface heights,
and consequently the probability to have an infinite peak and valley is non zero.  So an infinite pressure is needed for an infinite valley to reach the contact. 
As it was already mentioned, to obtain an analytical model comparable with real experiments, it is worth considering a truncated Gaussian surface as it is inevitably done in numerical models.}.
Nevertheless, the general trend predicted by Persson's model Eq.~\eqref{eq:area_persson} near the full contact is in a good agreement with the numerical results (Fig.~\ref{fig:near_full_contact}). 

For large nominal pressures $p_0$, Persson's prediction~\eqref{eq:area_persson} can be approximated
as 
\be \mbox{erf}\left(\frac{p_0}{E^*}\sqrt{\frac{2}{\langle|\nabla
    h|^2\rangle}}\right) \approx
1-\frac{\exp\left[-\left(\frac{p_0}{E^*}\sqrt{\frac{2}{\langle|\nabla
        h|^2\rangle}}\right)^\eta\right]}{\frac{p_0}{E^*}\sqrt{\frac{2\pi}{\langle|\nabla
      h|^2\rangle}}},\quad\eta=2.
\label{eq:full_area_approx}
\ee 
In our simulations, close to the full contact, the contact area converges to the full contact slightly faster than Persson's model predicts.
However, the exponent of approaching the full contact for our simulations is roughly the same as in Persson's model $\eta=2$ in~\eqref{eq:full_area_approx}. 
To demonstrate it, we plot the data as (see inset in Fig.~\ref{fig:near_full_contact})
\be
  F(p_0)=-\log\left[(1-A(p_0)/A_0)\frac{p_0}{E^*}\sqrt{\frac{2\pi}{\langle|\nabla h|^2\rangle}}\right].
  \label{eq:scaled_area}
\ee
Close to the full contact the numerical data points are grouped by their upper cutoff wavenumber $k_s$.
This is in contrast with the results for the rest of the contact areas (up to 80\%), where they are grouped by the lower cutoff $k_l$ (see Fig.~\ref{fig:area_num_1} and \ref{fig:area_slope_num}). We observe also that for a given $k_s$, the curve converges faster to the full contact for higher $\alpha$ (or smaller $k_l$).
This result should not be associated with numerical errors, as the discretization of the shortest wavelength $\lambda_s=2\pi/k_s$ is the same for each group of curves.
See Section~\ref{sec:num_error} for a discussion of inherent errors associated with application of numerical methods to problems of contact between rough surfaces.

\section{\label{sec:num_error}Corrected contact area and bounds on numerical errors}

The particularity of contact problems is that the contact pressure or
its derivative can diverge at the edge of the contact zone.  For example,
indentation of an elastic half-space by a rigid cylindrical stamp with radius $a$ along the axis of its symmetry
 leads to the contact pressure distribution $ p(r) \sim 1/\sqrt{1-r^2/a^2}$,
where $r$ is the distance from the center. So the contact pressure
as well as its derivative tend to infinity at the edge of the
contact zone $r=a$.  In the case of contact between a rigid sphere and an
elastic half-space the contact pressure is $ p(r) \sim \sqrt{1-r^2/a^2}$, 
where $a$ is the radius of the contact zone. 
The derivative of this contact pressure diverges at the contact edge.  
Thus it is difficult to estimate accurately the real contact area for a
discretized surface as the infinite slope cannot be captured by continuous 
interpolation functions. However, the precision may be
increased by the refinement of the discretization in the proximity of
the contact edges as suggested in~\cite{putignano2012ijss}. 
This refinement, however, cannot be implemented in arbitrary methods or may be computationally expensive. 
Another possibility is to use algorithms inspired from the level-set,
XFEM (extended finite element method)~\cite{xfem} or PUM (partition of unity method)~\cite{Melenk1996} methods,
which are in turn based on enriching the element interpolation functions~\cite{Heyliger1,Babuska1995}.
A promising application of these methods for an accurate tracking of the evolution of the edge of contact zone was recently reported in~\cite{chevaugeon2013conf}.

\begin{figure}[htb!]
\begin{center}
\includegraphics[width=0.7\textwidth]{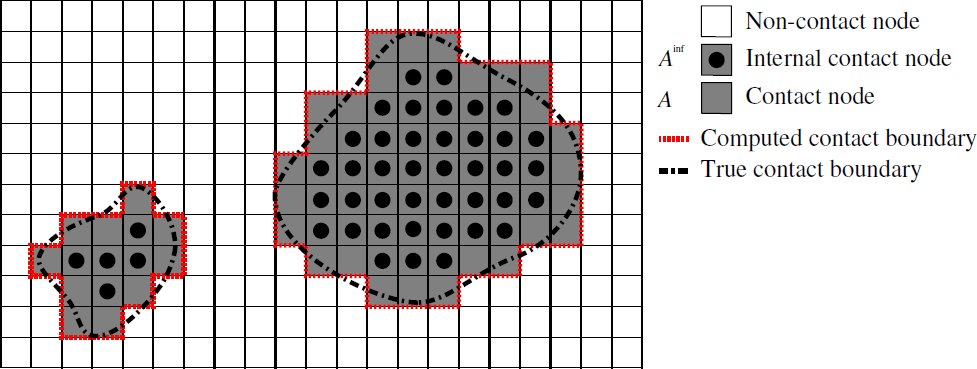}
\end{center}
\caption{\label{fig:error_graph}Example of a computed contact area
  (dark gray), internal nodes (black circles) that are in contact but
  are not adjacent to the contact boundary (red dashed line denotes the computed contact perimeter), and the boundary
  of the true contact area (dash-dotted line denotes the true contact perimeter).}
\end{figure}

\begin{figure}[htb!]
\includegraphics[width=1\textwidth]{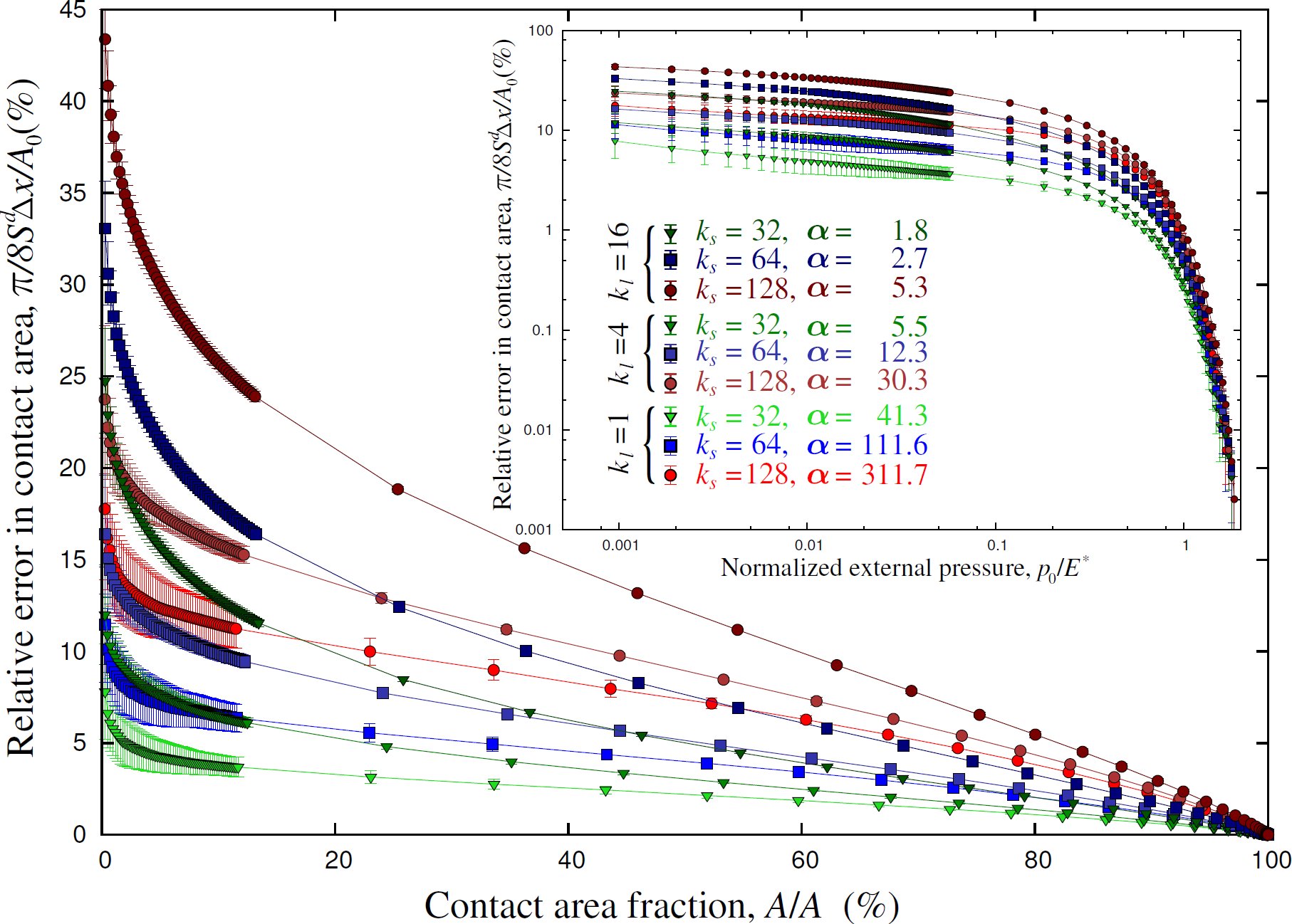}
\caption{\label{fig:relative_error}
  Approximation of the relative error in numerical estimation of the contact area as a function of the contact area fraction $A/A_0$ and as a
  function of the normalized contact pressure $p_0/E^*$ (in the inset in logarithmic scale) for rough surfaces discretized in 1024$\times$1024 points.}
\end{figure}

For a discretized surface, the true contact area fraction denoted by
$A^\prime/A_0$ is bounded between two values\footnote{It is
  true under the condition that the mechanical equations are
  accurately resolved for a given discretization. It implies that
  the discretization of the shortest wavelength in the surface spectrum is
  sufficiently fine.} (see Fig.~\ref{fig:error_graph}): (1) the relative
number of mesh nodes\footnote{We imply that a certain area is assigned to each node.} that do not lie on the edge of contact zones,
internal points, yielding $A^{\mbox{\tiny$\inf$}}/A_0$ and (2) all nodes that
are in contact, giving $A/A_0$.
$$ A^{\mbox{\tiny$\inf$}}/A_0 \quad\le\quad A^\prime/A_0 \quad\le\quad A/A_0. $$ 
These bounding values can be reliably estimated if in addition to
computing the contact area $A/A_0$ one computes the perimeter $S^d$ of
contact zones\footnote{To compute the relative contact area, we simply
  compute the number of nodes in contact and normalize it by the total
  number of nodes; to compute the perimeter, we compute the number of
  switches from contact to non-contact and \textit{vice versa} at each
  horizontal and vertical line of nodes.}. 
If the mesh spacing tends to zero the measure of the area fraction tends to a continuous limit, but it is not the case for the perimeter.
The discrete perimeter $S^d$ presents a perimeter of contact area measured using Manhattan (or $L1$) metric.
To obtain an Euclidean measure ($L2$) in a continuous limit one needs to correct the measured perimeter as $S=\pi/4\, S^d\;$\footnote{This multiplier $\pi/4$ can be easily obtained if we estimate the ratio in length between a hypotenuse and the sum of two cathetus for a rectangle triangle with angle $\phi$ and next we average this ratio over all $\phi\in[0;\pi/2]$.
Where the length of the hypotenuse represents a real perimeter and the sum of two cathetus its approximation via Manhattan metric. Another way to show it consists in computing
the perimeter  of a circle of radius $a$ using Manhattan metric, which gives $8a$, it is nothing but the perimeter of the smallest square in which this circle is inscribed. 
As the true contact perimeter is $2\pi a$, the ratio between the measured and the true perimeters is simply $4/\pi$.
}.
Then we estimate the lower bound of the contact area as
$$ A^{\mbox{\tiny$\inf$}} = \left(A-\frac{\pi}{4} S^d\Delta x\right),$$
where $\Delta x = L/N$ is the distance between nodes (mesh step)
and $N$ is the total number of nodes per side. Then the 
contact area can be roughly estimated as a mean value between the lower and the upper bounds
\be
  A'/A_0 \approx \left(A-\frac{\pi}{8} S^d\Delta x\right)/A_0,
\label{eq:corrected_area}
\ee
and the estimation of the relative error in numerical computation of the contact area for a given discretization is given by
$$ E(p_0)=\frac{\pi}{8} S^d\Delta x / A_0.$$
This rough estimation of the relative error is plotted in Fig.~\ref{fig:relative_error} as a function of the contact 
area and as a function of the contact pressure in logarithmic scale in the inset. 
This estimation may be as high as 45\%. One however has to bear in mind that the actual error may be significantly smaller (see Fig.~\ref{fig:error_graph}).
Even though this error decreases convexly, it remains high over the entire range of the contact area evolution: 
the error is only divided by two at about 30\% of the contact area fraction.
As this error estimation is proportional to the mesh step $\Delta x$, the error obtained in studies with finer meshes is proportionally smaller.
Predictably, surfaces with greater density of asperities suffer from higher relative error.
Also, surfaces with higher $k_l$ and/or $k_s$ have a higher relative error, as they generate more contact clusters.

\begin{figure}[htb!]
\includegraphics[width=1\textwidth]{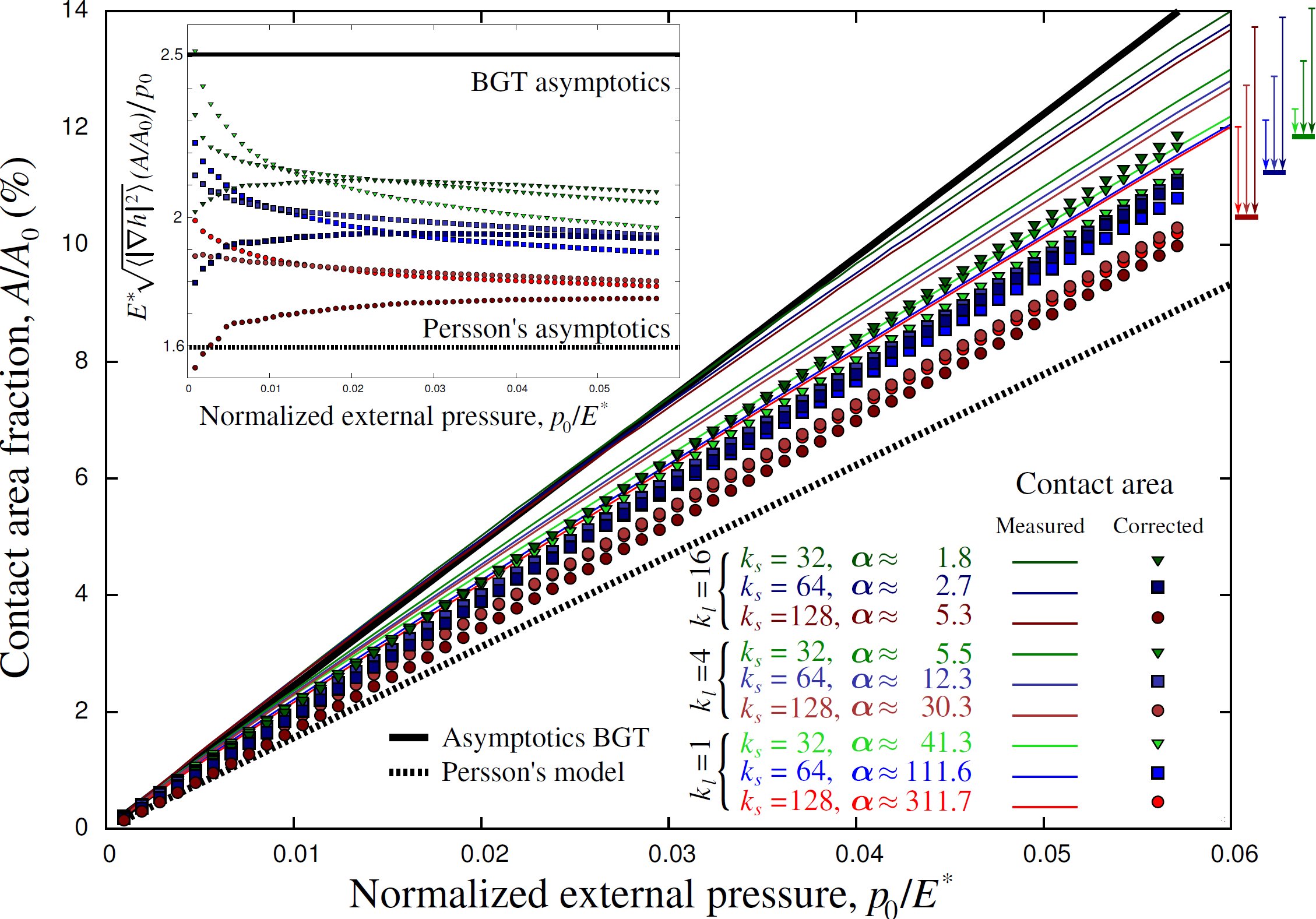}
\caption{\label{fig:corrected_area}
  Roughly corrected contact area (color points) and measured contact area (color lines, which correspond to points in Fig.~\ref{fig:area_num_1}) compared with Persson's model and BGT asymptotics. Color arrows on the right side depict approximately the shift of contact area values at $p_0=0.06 E^*$. In the inset the secant $E^*\sqrt{\langle|\nabla h|^2\rangle} (A/A_0)/p_0$ is plotted.}
\end{figure}

The approximately corrected contact area evolutions are depicted in Fig.~\ref{fig:corrected_area}.
These results are less accurate than the original data plotted in Fig.~\ref{fig:area_num_1}, but they may serve to indicate the trends and future work.
The corrected contact area~\eqref{eq:corrected_area} is significantly lower than the one measured directly and lies between the asymptotic predictions of Persson's and asperity based models.
The results depend only slightly on the lower cutoff wavenumber $k_l$, but strongly on the upper cutoff $k_s$.
The curves are grouped by the value of $k_s$; within each group there is no specific trend with respect to the lower cutoff or Nayak's parameter.
Also there is no consistency in the slope change (see inset in Fig.~\ref{fig:corrected_area}): for less Gaussian curves the secant evolves as a decreasing function,
whereas for the curves with $k_l=16$ it is an increasing function. It must be connected with a \emph{too rough estimation of the relative error}.
The estimation of $\kappa$ as a limit of secant as pressure tends to zero becomes ambiguous.
Although the correction of the contact area is quite rough, in future it may help to 
refine the results and to better carry out mesh convergence tests.

understand better how the Hurst exponent and cutoffs affect the evolution of the contact area. 
 
\section{\label{sec:disc}Discussion}

\subsection{Summary}
In this study we aimed to address how the \emph{true contact area} evolves from infinitesimally small to full contact for rough surfaces pressed together.
For that, we carried out a statistically meaningful numerical analysis using FFT based boundary element method for carefully synthesized rough surfaces.
We confirmed the key role of the lower cutoff wavenumber in the surface spectrum $k_l$ for the Gaussianity of generated random rough surfaces: 
surfaces with greater $k_l$ have distribution of heights close to a normal one~\cite{yastrebov2012pre},  
whereas surfaces with $k_l=1$ cannot be assumed normal even approximately (see Fig.~\ref{fig:normality} and \ref{fig:a:pdf_height}).
Thus for such surfaces the classical and widely used descriptions~\cite{longuethiggins1957rsla,nayak1971tasme} and 
associated analytical mechanical models~\cite{bush1975w,mccool1986w,thomas1999b,persson2001jcp,greenwood2006w,carbone2008jmps} cannot be applied.

All reported results are based on a relatively large number of simulations: for each combination of cutoff wavenumbers $k_s,k_l$ 
we performed simulations on 50 statistically equivalent realizations of rough surfaces.  
In combination with a relatively high number of data points at light contact (50 load steps up to $10-14\%$ of the contact area fraction),
it enables us to evaluate accurately derivatives of relevant characteristics (the contact area slope Fig.~\ref{fig:area_slope_num} and the derivative of the mean
contact pressure with respect to the applied load Fig.~\ref{fig:a:dpdp0}). 

\subsection{Effect of cutoff wavenumbers in the surface spectrum}
We studied the influence of the lower cutoff wavenumber, on the evolution of the contact area.
It was demonstrated that independently of the upper cutoff $k_s$, surfaces with greater value of $k_l$ (Gaussian surfaces) develop a bigger contact area fraction under a given contact pressure, than surfaces with low $k_l$ (non-Gaussian surfaces)\footnote{Remark that the root mean square gradient is kept constant for all considered surfaces independently on the cutoffs.}. This result holds up to about $A/A_0\approx80\%$.
Increasing $k_l$ from $1$ to $16$ does not lead to a convergence in the value of the contact area's slope, which continues to grow (see Fig.~\ref{fig:area_slope_num}),
even though for $k_l=16$ the surfaces may be assumed Gaussian with a high confidence.
We propose two arguments to explain this surprising behavior.
First, as shown in Section~\ref{sec:num_error}, the relative error in area estimation is significantly higher for surfaces with greater values of $k_l$.
An approximate account of this error in calculation of the contact area (Fig.~\ref{fig:corrected_area}) shifts all results to lower values lying between predictions of asperity based and Persson's models. This correction also makes them independent of the lower cutoff. On the other hand, the new results depend strongly on the upper cutoff $k_s$. No universality was found.
To confirm this assumption, a similar study has to be performed on different levels of discretization as in~\cite{campana2007epl,putignano2013ti}, 
or even better, an asymptotic dimensionless analysis can be done as in~\cite{prodanov2014tl}.
Second, the observed dependence of the upper (lower) cutoff wavenumbers may be assigned to varying density of asperities (see Eq.~\eqref{eq:D}) 
and to clustering of high asperities, which is quite pronounced for low $k_l$ (compare Fig.~\ref{fig:area_evolution_pic} and \ref{fig:area_evolution_pic2}).
The density of asperities controls the intensity of elastic interactions between contacting asperities resulting in different mechanical responses. 
Another study is required to confirm this assumption, it can be carried out using, for example, semi-analytical asperity based models that take into 
account the interaction between asperities~\cite{paggi2010w,yastrebov2011cras}.

\subsection{Nayak's parameter}
It is well known that Nayak's parameter $\alpha$\footnote{We recall that $\alpha\in[1.5;\infty]$ characterizes the breadth of the
surface spectrum, the higher $\alpha$, the more different wavelengths are contained in the surface.} 
influences significantly the statistical properties of rough surfaces~\cite{nayak1971tasme,greenwood2006w}.
In asperity based models, even at light pressures, the contact areas obtained for different $\alpha$ differ drastically~\cite{carbone2008jmps}; 
they coincide only at infinite separation (or infinitesimal contact areas) which cannot be accurately considered in simulations nor in real experiments.
An in-depth analysis of the role of $\alpha$ in asperity based models with and without interactions was conducted in~\cite{paggi2010w}.
However, in recent complete numerical studies\footnote{By a complete numerical study we imply an accurate resolution of mechanical boundary value problem subjected to contact constraints.} the Nayak's parameter was not discussed~\cite{hyun2004pre,pei2005jmps,campana2007epl,hyun2007ti,campana2008jpcm,campana2011jpcm,almqvist2011jmps,pohrt2012prl,putignano2012ijss,putignano2012jmps,yastrebov2012pre,putignano2013ti,prodanov2014tl}.
Many of these studies consider the effect of the upper cutoff wavenumber $k_s$ and the Hurst roughness exponent.
As $\alpha$ is strongly connected to these characteristics, this parameter was implicitly present in all forementioned studies, but was not discussed.
Here, we deduced simple formula~\eqref{eq:alpha} that links Nayak's parameter with $k_s, k_l$ and $H$ (see~\ref{app:alpha}\footnote{We also deduce a similar equation for the density of asperities, we also demonstrate that computed Nayak's parameter and density of asperities are in very good agreement with predictions of the proposed formulae.}), that will allow to put 
the previous numerical results in position to reveal the role of $\alpha$.

In this preliminary study we demonstrated that (see Fig.~\ref{fig:area_num_1},\ref{fig:area_slope_num}): 
(i) for a given pressure and a fixed lower cutoff $k_l$, a greater $\alpha$ results in a lower contact area, which is in qualitative agreement with asperity based models~\cite{greenwood2006w,carbone2008jmps}; 
(ii) however, the effect of $\alpha$ is exaggerated in these models, as we obtain a considerably smaller decrease in the contact area for a wide range of $\alpha$ values.
After studying surfaces with different Hurst exponents and different Nayak's parameter, 
we conjecture that the Hurst roughness exponent does not play an independent role in mechanical behavior of rough
surfaces, but only via the Nayak's parameter and cutoff frequencies, as they are related via Eq.~\eqref{eq:a:alpha}. 
Thus it would be important in future to compare rigorously the evolution of the contact area for different $H$ and similar $\alpha$ or \textit{vice versa}.

\subsection{Coefficient of proportionality $\kappa$ between normalized area and pressure}
Regardless of our efforts in obtaining statistically meaningfull (every data point is an average over $50$ surfaces) 
and accurate results at light contact ($50$ load steps up to $10-14\%$ of contact area), the theoretical 
coefficient of proportionality $\kappa$ between the contact area fraction and the normalized pressure
cannot be accurately estimated (see Fig.~\ref{fig:area_slope_num}). 
This is because the slope of the contact area changes rapidly at light pressure.
Even if one possesses a reliable analytical expression for the contact area evolution, the range of fittable $\kappa$ is rather wide.
Moreover, at very small contact areas (relevant to estimation of $\kappa$), 
the upper bound of the relative error may be huge (see Fig.~\ref{fig:relative_error}), that prevents obtaining reliable results in all numerical simulations of this type.
Only a rough estimation of $\kappa$ can be given:  
it is close to the prediction of asperity based models $\kappa_{BGT}=\sqrt{2\pi}$, and must be higher than predicted by Persson's model $\kappa_{P}=\sqrt{8/\pi}$.
The corrected areas in Fig.~\ref{fig:corrected_area} suggest significantly lower values for $\kappa$.
But as the error estimation remains quite approximate, this result should be considered prudently.
In general, we suggest to abandon the consideration of this equivocal (or purely mathematical) coefficient 
and focus on the nonlinear evolution of the contact area up to 10-20\%.

\subsection{Contact evolution formula}
The phenomenological contact evolution formula~\eqref{eq:cel} that we suggested in~\cite{yastrebov2012pre}, 
whose derivation is given in~\ref{app:cel}, fits well all our results at small contact area fractions ($<20\%$). 
The fit parameters, however, cannot be assumed universal.
It is clear that some normalization parameters are missed, 
and introducing explicitly the asperity density in equations may yield a more universal equation for the
contact area evolution.

\subsection{General trends in contact area evolution}
An important and reliable result that we obtained is that the
derivative of the contact area with respect to the contact pressure is
a decreasing convex function.  This result is in a good agreement with
asperity based models (see, for example,~\cite{bush1975w,carbone2008jmps}) 
and contradicts Persson's model~\cite{persson2001jcp}, which predicts concavity (see inset in Fig.~\ref{fig:area_evolution_full}). 
We suppose that this prediction of the Persson's model may arise from 
(i) the artificial~\cite{manners2006w} extension of the model from the full contact (at which the model is accurate) to partial contact~\cite{persson2002prb} and 
(ii) from not accounting for Nayak's parameter in deducing the associated Persson's diffusion equation.
Consideration of a varying mean curvature of asperities at different heights may probably improve this model for small Nayak's parameter. 
Still it should not have an effect for great values of $\alpha$, for which the mean curvature is almost independent of height~\cite{nayak1971tasme}.
Thus Persson's model may be considered only for surfaces with high $\alpha$, that is often the case for real surfaces.
Predictably, the corrected areas for greater $\alpha$ and for a given $k_l$ are closer to Persson's prediction (see Fig.~\ref{fig:corrected_area}). 
We also have to bear in mind that all analytical models are deduced for surfaces with Gaussian distribution of heights, 
so they can be reliably compared only with numerical results obtained  for high values of $k_l$, which yield Gaussian surfaces. 

\section{Conclusions}
We consider the breadth of the surface spectrum (Nayak's parameter $\alpha$) in numerical simulation of rough  frictionless and non-adhesive contact between linear elastic solids with self-affine roughness.
For that we deduce a simple formula linking the Hurst exponent and cutoffs in the surface spectrum.
We demonstrate that to generate a periodic Gaussian surface the lower cutoff wavenumber has to be high enough, 
i.e. the shortest wavelength has to be significantly smaller than the period of the surface.
We introduce bounds on numerical error in estimation of the contact area in numerical models.
All these results suggest that the contact area evolution is non-linear and dependent on Nayak's parameter.
This results are in good qualitative agreement with elaborated asperity based models.
On the other hand the obtained results show that at light pressures Persson's model should be modified, 
as it predicts a quasi-linear and universal evolution of the contact area independently of Nayak's parameter.

Within the suggested rigorous framework and the computational power of modern computers and software of research groups involved in this domain, 
we hope that numerous unresolved questions will be successfully addressed in prospective studies.
The contact area controls many physical, mechanical and tribological phenomena at the interface between contacting solids: friction, wear, adhesion, fretting and energy transfer.
Thus, understanding the mechanics of contact and precisely the evolution of the contact area is an important issue for many applications: electrical contact, protective coatings,
and many others.

\section{Acknowledgment}

GA and JFM greatly acknowledge the financial support from the European Research Council (ERCstg UFO-240332).

\appendix
\section{\label{app:alpha}Calculation of Nayak's parameter and density of asperities}

\subsection{Nayak's parameter}

\begin{figure}
 \begin{center}
  \includegraphics[width=0.8\textwidth]{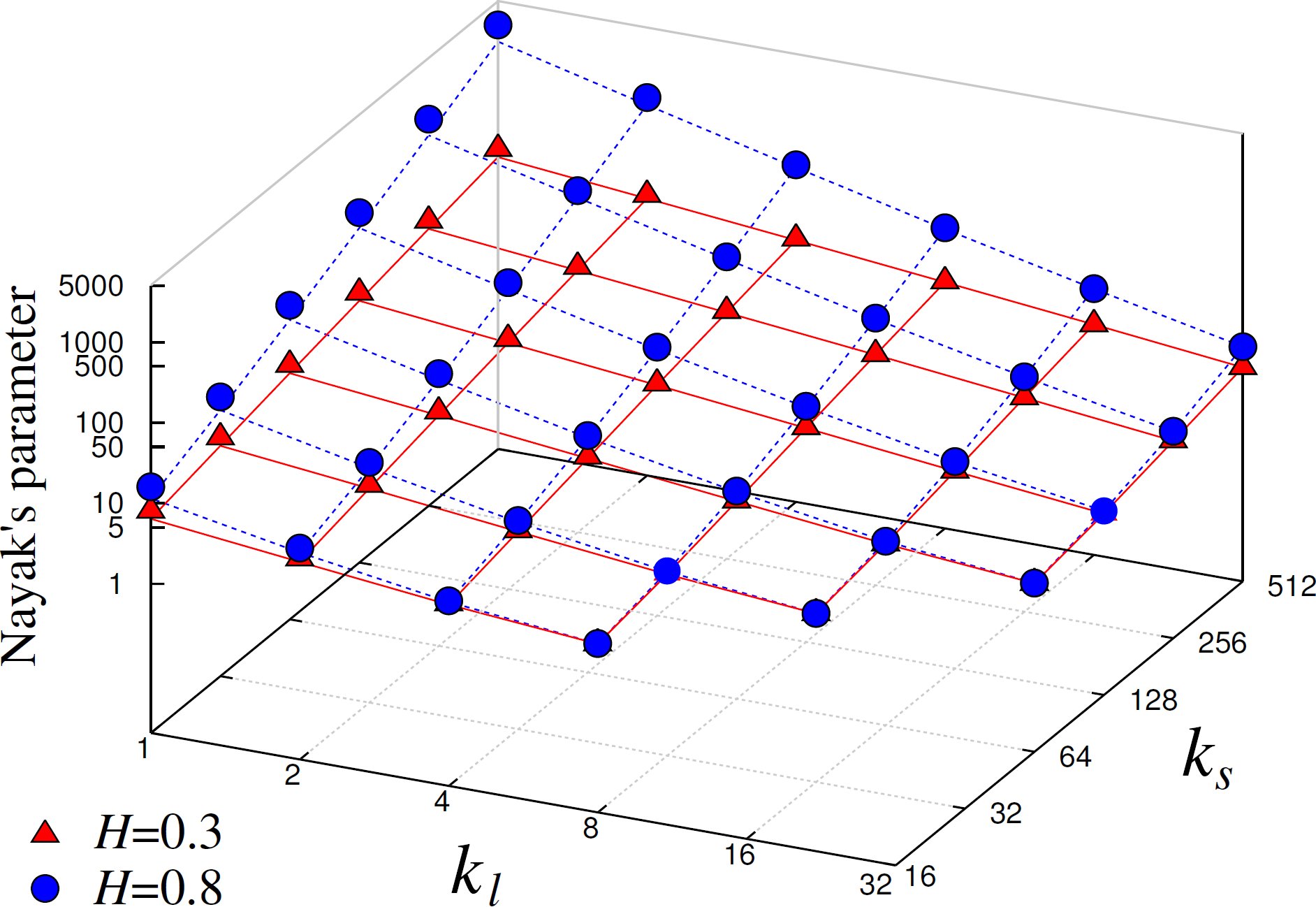}
 \end{center}
 \caption{\label{fig:a:1}Computed Nayak's parameter for different
   cutoffs and two Hurst exponents ($H=0.3$ circles and $H=0.8$
   triangles); every value is an average over 200 surfaces and
   compared with analytical predictions~\eqref{eq:a:alpha} ($H=0.3$
   red solid and $H=0.8$ blue dashed lines).}
\end{figure}

The Nayak's parameter $\alpha$ that characterizes the breadth of the
surface spectrum has a paramount effect on the properties of rough
surfaces~\cite{nayak1971tasme}: mean and principal curvatures of
asperities, relations between asperity's height and curvature,
anisotropy of asperities, probability density of asperity's heights,
all depend on $\alpha$. Consequently it plays a major role in
mechanics of rough contact. The Nayak's parameter for an isotropic surface is computed as the
following ratio between 3 moments of the power spectral density (PSD)
$\Phi^s(k_x,k_y)$ of the surface: \be
\alpha=\frac{m_{00}\,m_{40}}{m_{20}^2},
\label{eq:a:a1}
\ee where
$$ m_{pq} = \iint\limits_{-\infty}^{\quad\infty} k_x^pk_y^q
\Phi^s(k_x,k_y) dk_xdk_y.
$$ 
We assume that the PSD is axisymmetric (depends only on the norm of the wavevector $|\mathbf k|=\sqrt{k_x^2+k_y^2}$) and is bounded between wavenumbers $k_l$ and $k_s$: 
\bes \Phi^s(|\mathbf k|)
= \begin{cases} 
  Ak_r^{-2(1+H)},&\mbox{ if } k_l \le |\mathbf k| < k_r;\\ 
  A|\mathbf k|^{-2(1+H)},&\mbox{ if } k_r \le |\mathbf k| \le k_s;\\
  0,&\mbox{else,}
            \end{cases}
\ees 
where $k_l$ is the lower cutoff wavenumber in the surface
spectrum, next the PSD does not change up to $k_r$ (also referred as a
roll-off), at which occurs the transition from the plateau frequency zone to
the power law decrease, and $k_s$ is the upper cutoff wavenumber (see Fig.~\ref{fig:a:k_lks}).
In the paper we consider surfaces without plateau, so $k_r=k_l$, but for the sake of generality we
derive the complete expressions.  The $m_{q0}$ moment is given by
$$ m_{q0} = \iint\limits_{-\infty}^{\quad\infty} k_x^q \Phi^s(k_x,k_y)dk_xdk_y.$$ 
Due to the symmetry of the PSD and the introduced cutoff wavenumbers we can replace this integral by
\be
\begin{split}
&m_{q0}=\int\limits_{k_l}^{k_s}\int\limits_{0}^{2\pi} \left[k
    \cos(\varphi)\right]^q \Phi^s(k_x,k_y) dk_xdk_y=
  A\int\limits_{k_l}^{k_r}\int\limits_{0}^{2\pi} \left[k
    \cos(\varphi)\right]^q k_r^{-2(1+H)}\,kdkd\varphi +\\&+
  A\int\limits_{k_r}^{k_s}\int\limits_{0}^{2\pi} \left[k
    \cos(\varphi)\right]^qk^{-2(H+1)} k dkd\varphi =
  A\int\limits_{0}^{2\pi} \cos^q(\varphi)d\varphi\left(
  k_r^{-2(1+H)}\int\limits_{k_l}^{k_r} k^{q+1} dk +
  \int\limits_{k_r}^{k_s} k^{q-2H-1} dk\right) =\\ &=
  AT(q)\left(k_r^{-2(1+H)}\left.\frac{k^{q+2}}{q+2}\right|_{k_l}^{k_r}+\left.\frac{k^{q-2H}}{q-2H}\right|_{k_r}^{k_s}\right)
  =
  AT(q)k_r^{q-2H}\left(\frac{1-\xi^{q+2}}{q+2}+\frac{\zeta^{q-2H}-1}{q-2H}\right),
\end{split}
\label{eq:a:mq0}
\ee
where $\xi=k_l/k_r$, $\zeta = k_s/k_r$ ($\zeta$ is referred as magnification in Persson's model~\cite{persson2001jcp}), and $T(q)$
\be
T(q) = \int\limits_{0}^{2\pi} \cos^q(\varphi)d\varphi
= \begin{cases} 2\pi,&\mbox{ if }q=0;\\ \pi,&\mbox{ if
  }q=2;\\ 3\pi/4,&\mbox{ if }q=4.\\
 \end{cases}
 \label{eq:a:cq}
\ee 
Thus, the Nayak's parameter~\eqref{eq:a:a1} is \be \boxed{
  \alpha(H,\zeta,\xi) =\frac{3}{2}
  \left.\left(1-\xi^2+\frac{1-\zeta^{-2H}}{H}\right)\left(\frac{1-\xi^6}{3}+\frac{\zeta^{4-2H}-1}{2-H}\right)
  \;\right/\;
  \left(\frac{1-\xi^4}{2}+\frac{\zeta^{2-2H}-1}{1-H}\right)^2 }
\label{eq:a:gen_alpha} 
\ee In this paper we generate rough surfaces without plateau, so
$k_r=k_l$, $\xi=1$, $\zeta=k_s/k_l$ and Eq.~\eqref{eq:a:gen_alpha} reduces to 
\be \boxed{
  \alpha(H,\zeta) =\frac{3}{2}
  \frac{(1-H)^2}{H(H-2)}\frac{(\zeta^{-2H}-1)(\zeta^{4-2H}-1)}{(\zeta^{2-2H}-1)^2}
}
\label{eq:a:alpha} 
\ee 
The asymptotics of $\alpha$ for high $\zeta$ is given by 
\be
\boxed{\alpha \sim \zeta^{2H}}
\label{eq:a:alpha_as}
\ee Note that this function changes the convexity at $H=0.5$ (concave
for $H<0.5$, convex for $H>0.5$).  Thus, the value $H=0.5$ can be
considered as a critical roughness exponent.
Note also that Nayak's parameter depends only on the ratios $\xi=k_r/k_l$,
$\zeta=k_s/k_r$ and the Hurst exponent.  To validate the
formula~\eqref{eq:a:alpha} for surfaces without a plateau, for each combinations of cutoff
wavenumbers $k_l=k_r=\{1,2,4,8,16,32\}$,
$k_s=\{16,32,64,128,256,512\}$, we compute the Nayak's
parameter for 200 generated surfaces $1024\times1024$, report the mean value and
compare it to the formula~\eqref{eq:a:alpha} (see Table~\ref{tab:a:1} and
Fig.~\ref{fig:a:1}).  Real and analytical results are in good
agreement, except that at low $k_l$ the analytical predictions
underestimates considerably the computed results as such surfaces cannot be considered isotropic and Gaussian.
For high $k_l$ we find a quasi perfect agreement between the formula and the computed value.

\begin{table}
\begin{center}
\begin{tabular}{c|cccccc}
  \multicolumn{7}{l}{$H=0.3$ numerical}\\ $k_l=k_r\setminus k_s$ & 16
  & 32 & 64 & 128 & 256 & 512\\ \hline 1 & 8.06 & 12.91 & 20.31 &
  31.49 & 48.46 & 74.20 \\ 2 & 4.39 & 7.11 & 11.34 & 17.80 & 27.66 &
  42.64 \\ 4 & 2.60 & 4.14 & 6.68 & 10.64 & 16.73 & 26.02\\ 8 & 1.76 &
  2.56 & 4.06 & 6.53 & 10.41 & 16.38\\ 16 & - & 1.75 & 2.54 & 4.01 &
  6.45 & 10.28\\ 32 & - & - & 1.75 & 2.53 & 4.00 & 6.43\\
\end{tabular}
\begin{tabular}{c|cccccc}
  \multicolumn{7}{l}{$H=0.3$ analytical, Eq.~\eqref{eq:a:alpha}}\\ $k_l=k_r\setminus k_s$ & 16
  & 32 & 64 & 128 & 256 & 512\\ \hline 1 & 6.43 & 10.25 & 16.13 &
  25.10 & 38.74 & 59.43\\ 2 & 4.00 & 6.43 & 10.25 & 16.13 & 25.10 &
  38.74\\ 4 & 2.53 & 4.00 & 6.43 & 10.25 & 16.13 & 25.10\\ 8 & 1.74 &
  2.53 & 4.00 & 6.43 & 10.25 & 16.13\\ 16 & - & 1.74 & 2.53 & 4.00 &
  6.43 & 10.25\\ 32 & - & - & 1.74 & 2.53 & 4.00 & 6.43\\
\end{tabular}

\begin{tabular}{c|cccccc}
  \multicolumn{7}{l}{$H=0.8$ numerical}\\ $k_l=k_r\setminus k_s$ & 16
  & 32 & 64 & 128 & 256 & 512\\ \hline 1 & 16.05 & 41.31 & 111.61 &
  311.73 & 890.20 & 2583.14 \\ 2 & 5.94 & 13.60 & 33.91 & 89.57 &
  245.87 & 693.45 \\ 4 & 2.81 & 5.49 & 12.31 & 30.31 & 79.31 &
  216.31\\ 8 & 1.77 & 2.77 & 5.36 & 11.94 & 29.23 & 76.22\\ 16 & - &
  1.76 & 2.74 & 5.26 & 11.68 & 28.53\\ 32 & - & - & 1.76 & 2.73 & 5.23
  & 11.62 \\
\end{tabular}
\begin{tabular}{c|cccccc}
  \multicolumn{7}{l}{$H=0.8$ analytical, Eq.~\eqref{eq:a:alpha}}\\ $k_l=k_r\setminus k_s$ & 16
  & 32 & 64 & 128 & 256 & 512\\ \hline 1 & 11.60 & 28.33 & 73.73 &
  200.38 & 561.14 & 1604.91\\ 2 & 5.23 & 11.60 & 28.33 & 73.73 &
  200.38 & 561.14\\ 4 & 2.72 & 5.23 & 11.60 & 28.33 & 73.73 &
  200.38\\ 8 & 1.76 & 2.72 & 5.23 & 11.60 & 28.33 & 73.73\\ 16 & - &
  1.76 & 2.72 & 5.23 & 11.60 & 28.33\\ 32 & - & - & 1.76 & 2.72 & 5.23
  & 11.60\\
\end{tabular}

\end{center}
\caption{\label{tab:a:1}Computed Nayak's parameter for different
  cutoffs and two Hurst exponents $H=0.3, 0.8$ and analytical results
  Eq.~\eqref{eq:a:alpha}; numerical results are averaged over 200
  surfaces.}
\end{table}

\subsection{Density of asperities}

Similarly we can express the density of asperities through the cutoff frequencies and the Hurst exponent.
The average density of asperities as computed in~\cite{longuethiggins1957rsla,nayak1971tasme}:
\be
D=\frac{\sqrt{3}}{18\pi}\frac{m_4}{m_2}.
\label{eq:Danal}
\ee
So substituting \eqref{eq:a:mq0} and \eqref{eq:a:cq} in this formula we obtain the following expression:
\be
   \boxed{D(H,\zeta,\xi)=\frac{\sqrt3}{24\pi}k_r^2\left.\left(\frac{1-\xi^{6}}{3}+\frac{\zeta^{4-2H}-1}{2-H}\right)\;\right/\;\left(\frac{1-\xi^{4}}{2}+\frac{\zeta^{2-2H}-1}{1-H}\right)}
   \label{eq:D1}
\ee
If $k_r=k_l$ we obtain a simpler form relevant to our paper:
\be
   D(H,\zeta)=\frac{\sqrt3}{24\pi}\frac{1-H}{2-H}\frac{\zeta^{4-2H}-1}{\zeta^{2-2H}-1}k_l^2.
   \label{eq:D2}
\ee
When the surface contains many modes $\zeta\to\infty$ we may approximate the density of asperities as 
$$D\approx\frac{\sqrt3}{24\pi}\frac{1-H}{2-H}k_s^2.$$
From the density of asperities, we can easily compute the average distance between closest asperities:
$$
  \langle d\rangle = 1/\sqrt{D} \sim 1/k_s. 
$$ 
As close asperities have a comparable height, the critical contact radius for asperities, 
at which the associated contact zones coalesce, may be estimated as $a_c = \langle d\rangle/2$.

\begin{figure}
 \begin{center}
  \includegraphics[width=0.8\textwidth]{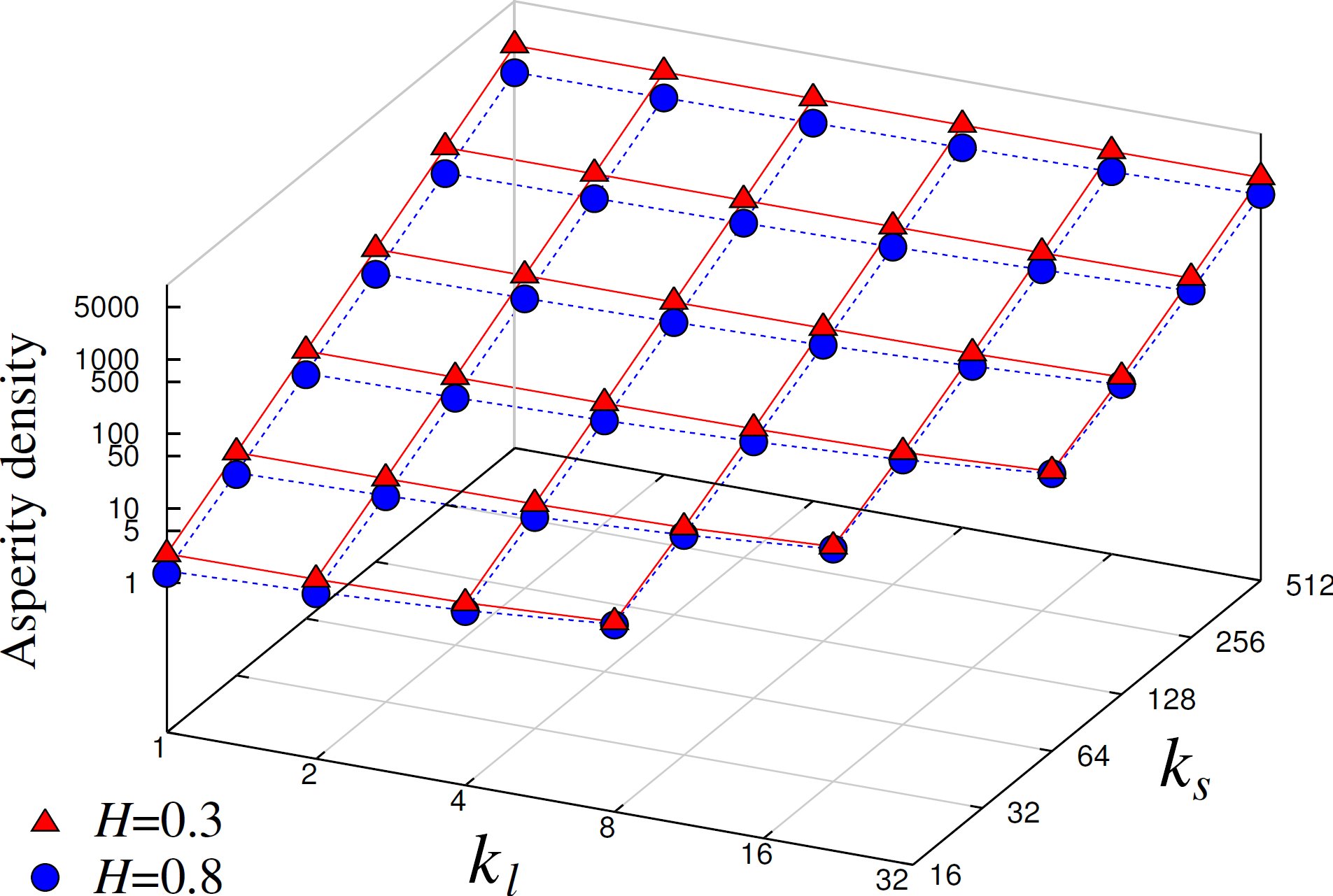}
 \end{center}
 \caption{\label{fig:a:1b}Computed density of asperities for different
   cutoffs and two Hurst exponents ($H=0.3$ circles and $H=0.8$
   triangles); every value is an average over 200 surfaces and
   compared with analytical predictions~\eqref{eq:D2} ($H=0.3$
   red solid and $H=0.8$ blue dashed lines).}
\end{figure}

As previously, for each combinations of cutoff
wavenumbers $k_l=k_r=\{1, 2, 4, 8, 16, 32\}$, $k_s=\{16, 32, 64, 128, 256, 512\}$, we compute directly the densities of asperities~\eqref{eq:Danal}
for  200 generated surfaces $1024\times1024$, evaluate its mean value and
compare it to the analytical prediction~\eqref{eq:D2} (see Table~\ref{tab:a:1b} and
Fig.~\ref{fig:a:1b}).  Real results and analytical predictions are in quasi perfect
agreement for all cutoffs, except for $k_l=1,2$ at which the analytical formula overestimates the real value by several percent.

\begin{table}
\begin{center}
\begin{tabular}{c|cccccc}
  \multicolumn{7}{l}{$H=0.3$ numerical}\\ 
  $k_l=k_r\setminus k_s$ & 16 & 32 & 64 & 128 & 256 & 512\\ \hline 
1 & 2.42  & 9.69  & 38.75  & 154.98  & 619.92  & 2479.72 \\
2 & 2.50  & 9.81  & 38.93  & 155.25  & 620.33  & 2480.35 \\
4 & 2.74  & 10.16  & 39.44  & 156.01  & 621.48  & 2482.09 \\
8 & 3.46  & 11.13  & 40.83  & 158.05  & 624.54  & 2486.69 \\
16 & -  & 14.06  & 44.76  & 163.63  & 632.72  & 2498.94 \\
32 & -  & -  & 56.41  & 179.21  & 654.85  & 2531.45 \\
\end{tabular}
\begin{tabular}{c|cccccc}
  \multicolumn{7}{l}{$H=0.3$ analytical, Eq.~\eqref{eq:D2}}\\
  $k_l=k_r\setminus k_s$ & 16 & 32 & 64 & 128 & 256 & 512\\ \hline 
1 & 2.47  & 9.76  & 38.86  & 155.15  & 620.17  & 2480.04 \\
2 & 2.56  & 9.89  & 39.05  & 155.44  & 620.61  & 2480.69 \\
4 & 2.80  & 10.23  & 39.56  & 156.20  & 621.75  & 2482.42 \\
8 & 3.53  & 11.21  & 40.94  & 158.23  & 624.79  & 2487.00 \\
16 & -  & 14.12  & 44.83  & 163.76  & 632.91  & 2499.15 \\
32 & -  & -  & 56.47  & 179.34  & 655.02  & 2531.64 \\
\end{tabular}

\begin{tabular}{c|cccccc}
  \multicolumn{7}{l}{$H=0.8$ numerical}\\ 
  $k_l=k_r\setminus k_s$ & 16 & 32 & 64 & 128 & 256 & 512\\ \hline 
1 & 1.35  & 4.96  & 18.64  & 71.36  & 276.27  & 1078.71 \\
2 & 1.64  & 5.65  & 20.43  & 76.19  & 289.75  & 1117.16 \\
4 & 2.15  & 6.76  & 23.07  & 82.92  & 307.76  & 1167.08 \\
8 & 3.21  & 8.77  & 27.36  & 93.04  & 333.40  & 1235.44 \\
16 & -  & 13.05  & 35.41  & 110.14  & 373.57  & 1336.97 \\
32 & -  & -  & 52.41  & 141.97  & 441.09  & 1495.36 \\
\end{tabular}
\begin{tabular}{c|cccccc}
  \multicolumn{7}{l}{$H=0.8$ analytical, Eq.~\eqref{eq:D2}}\\
  $k_l=k_r\setminus k_s$ & 16 & 32 & 64 & 128 & 256 & 512\\ \hline 
1 & 1.46  & 5.23  & 19.35  & 73.25  & 281.55  & 1093.87 \\
2 & 1.72  & 5.84  & 20.90  & 77.39  & 292.98  & 1126.22 \\
4 & 2.22  & 6.90  & 23.37  & 83.62  & 309.55  & 1171.93 \\
8 & 3.28  & 8.88  & 27.58  & 93.49  & 334.47  & 1238.22 \\
16 & -  & 13.12  & 35.52  & 110.32  & 373.95  & 1337.89 \\
32 & -  & -  & 52.49  & 142.08  & 441.29  & 1495.80 \\
\end{tabular}
\end{center}
\caption{\label{tab:a:1b}Computed density of asperities for different
  cutoffs and two Hurst exponents $H=0.3, 0.8$ and analytical predictions
  Eq.~\eqref{eq:D2}; numerical results are averaged over 200 surfaces.}
\end{table}

\section{\label{app:height_distr}Height distribution}

\begin{figure}[ht!]
 \begin{center}
 \includegraphics[width=.92\textwidth]{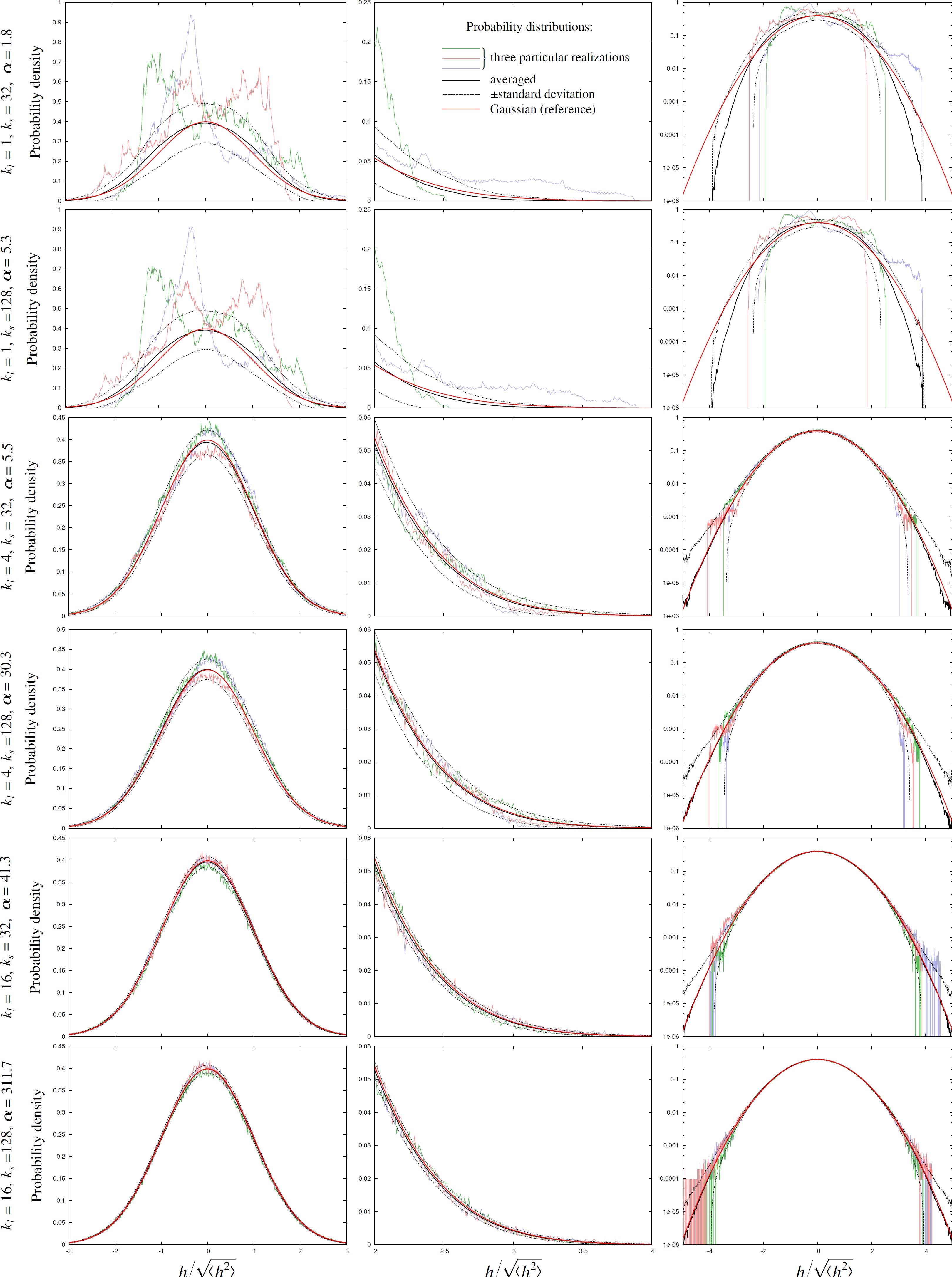}
 \caption{\label{fig:a:pdf_height}Surface heights distribution for
   different cutoff wavenumbers $k_l=1,4,16$, $k_s=32,128$ in linear (left and central columns) and semi-logarithmic plots (right column);
   in central column we give a zoom on the height distribution in $h/\sqrt{\langle h^2\rangle} \in [2;4]$.
   We plot height distribution for 3 particular surfaces (colored dashed oscillating lines), for a distribution averaged over 1000 statistically equivalent surfaces $H=0.8$ (solid black line),
   its standard deviation (dashed black line) and a reference Gaussian distribution (red solid line).}
  \end{center}
\end{figure}

Here we provide the reader with a more detailed plot of the height distribution (Fig.~\ref{fig:a:pdf_height}) overaged over 1000 surfaces for each combination of cutoffs.
One can see that increasing $k_l$ from $1$ to $4$ improves drastically the probability density of heights of particular surfaces 
(height distribution for three randomly chosen surfaces is plotted for each combination of cutoff wavenumbers), so that it approaches a normal distribution.
On the other hand an increase in $k_s$ does not result in approaching a normal distribution. 
One can also note that even averaged over 1000 surfaces, the averaged height distribution for $k_l=1$ is non-Gaussian. 
The standard deviation of the probability density is a good indicator of Gaussianity for particular
surface realizations within a given set of parameters; the standard deviation decreases rapidly with increasing $k_l$.

\section{\label{app:cel}Derivation of the contact evolution law}

\begin{figure}
 \begin{center}
  \includegraphics[width=1\textwidth]{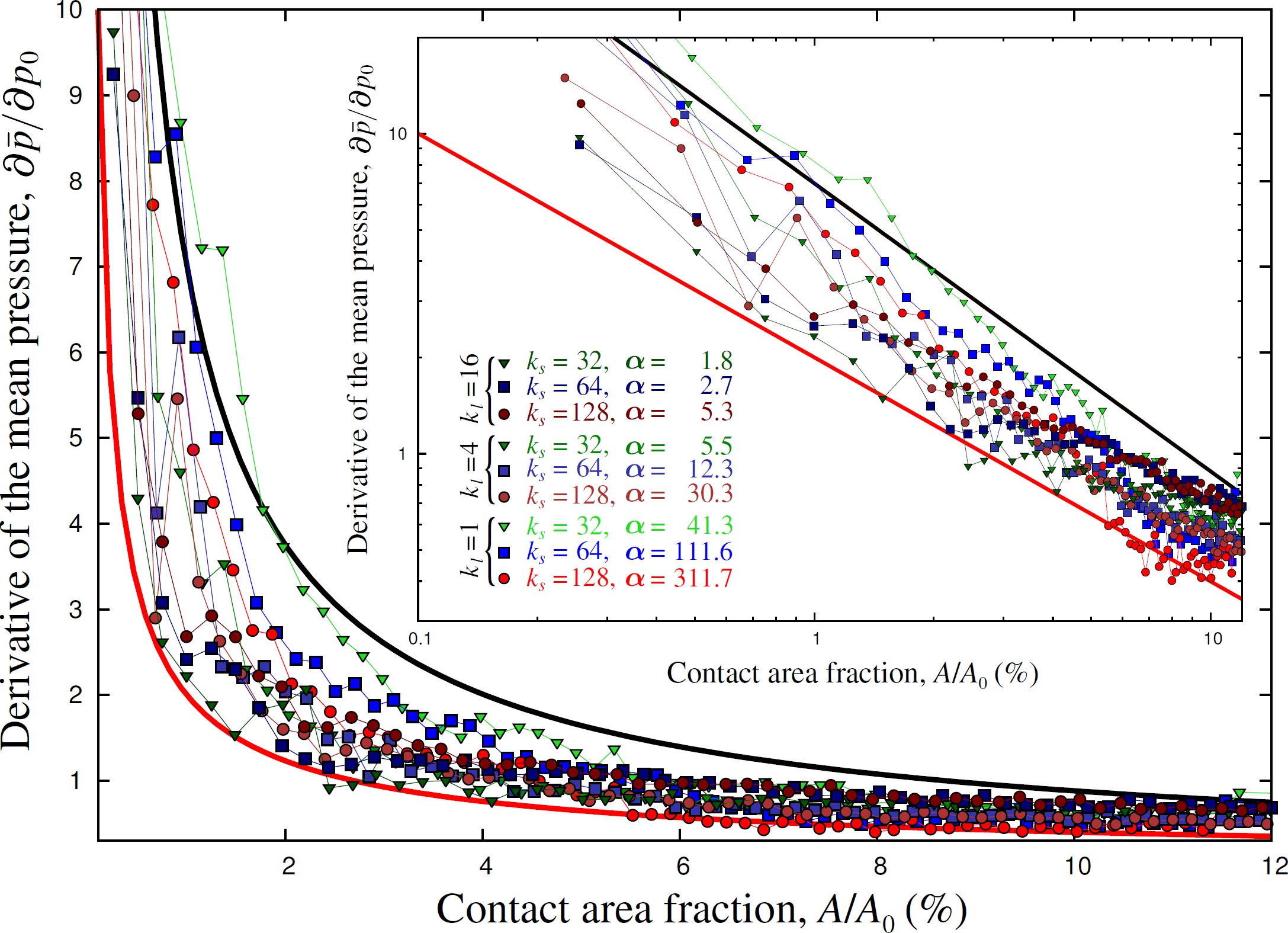}
 \end{center}
 \caption{\label{fig:a:dpdp0}Derivative of the mean contact pressure $\bar p=F/A$
   with respect to the external pressure $p_0=F/A_0$ and its evolution
   with the contact-area fraction $A/A_0$ for different cutoff
   wavenumbers; red line corresponds to the exponent $-0.7$, black one to $-0.9$.}
\end{figure}

In \cite{yastrebov2012pre} we demonstrated that the derivative of the mean contact pressure
 $\bar p = F/A$ with respect to the external pressure $p_0=F/A_0$ decreases approximately as a power law of the contact area:
\begin{equation}
 \label{eq:b1}
 \frac{d\bar p}{d p_0} = \beta\left(\frac{A_0}{A}\right)^{1-\mu},\; \beta
 > 0,\;0 < \mu < 1.
\end{equation}
By expanding the terms $\bar p$ and $p_0$ we obtain the equation
\[
 F\frac{d A}{d F} = A - \beta A_0\left(\frac{A_0}{A}\right)^{-\mu-1},
\]
which we integrate from $A_c$ to $A$ and from $F_c$ to $F$
\begin{equation}
\label{eq:a3} 
\int\limits_{A_c}^A\frac{d (A/A_0)}{A/A_0 - \beta
  \left(A/A_0\right)^{1+\mu}} = \int\limits_{F_c}^{F} \frac{dF}{F}
\quad\Rightarrow\quad
-\frac{1}{\mu}\ln\left(\frac{\left(A/A_0\right)^{-\mu} - \beta}{
  \left(A_c/A_0\right)^{-\mu} -\beta}\right) = \ln(F/F_c).
\end{equation} 
For the inferior integration limits $F_c \to 0,\;A_c \to 0$, we 
assume that the contact area is proportional to the pressure as in BGT model~\cite{bush1975w} (see Eq.~\eqref{eq:linear_a}), that we substitute in the left
hand side of equation~\eqref{eq:a3}:
\[
  A_c = \frac{\kappa}{\sqrt{\langle|\nabla
      h|^2\rangle\vphantom{A^{b}}} E^*} F_c \quad\Rightarrow\quad
  -\frac{1}{\mu}\ln\left(\frac{\left(A/A_0\right)^{-\mu} - \beta}{
    \left(\frac{\kappa}{\sqrt{\langle|\nabla h|^2\rangle} E^*}
    F_c/A_0\right)^{-\mu} -\beta}\right) = \ln(F/F_c).
\]
Since $\mu > 0$ and $F_c \to 0$, we can neglect the coefficient $\beta$ in the denominator;
by taking the exponential of the two sides of this equation, we obtain 
an equation for the evolution of real contact area with three parameters $\beta,\mu,\kappa$:
\begin{equation}
 \frac{\left(A/A_0\right)^{-\mu} - \beta}{
   \left(\frac{\kappa}{\sqrt{\langle|\nabla h|^2\rangle} E^*}
   F_c/A_0\right)^{-\mu}} = (F/F_c)^{-\mu} \quad\Rightarrow\quad
\boxed{ \frac{A}{A_0} = \frac{1}{\left[\beta + \left(\frac{\sqrt{\langle|\nabla h|^2\rangle}E^*}{\kappa p_0}\right)^{\mu}\right]^{1/\mu}}}
\label{eq:f}
\end{equation}


\end{document}